\documentclass{tlp}
\usepackage{amssymb}
\usepackage{graphicx}
\usepackage{amsmath}

\newtheorem{definition}{Definition}[section]
\newtheorem{theorem}{Theorem}[section]

\newtheorem{example}{Example}[section]
\newtheorem{lemma}{Lemma}[section]

\newtheorem{property}{Property}[section]
\input{tcilatex}

\begin{document}

\title[Redundancy Elimination Tolerant Scheduling Rules]{On Redundancy Elimination Tolerant\\
Scheduling Rules\thanks{%
This work has been partially supported by the Italian National Research
Council (CNR) research project ``Tecniche di taglio dei cicli e loro
implementazione in ambiente di Programmazione Logica'', Grant No.
97.02432.CT12.}}
\author[F. Ferrucci, G. Pacini and M.I. Sessa]
{FILOMENA FERRUCCI\\
DMI Universita' di Salerno - via S. Allende, 84081 Baronissi (SA), Italy\\
\email{filfer@unisa.it}
\and
 GIULIANO PACINI\\ 
Accademia Navale - Viale Italia 72, 57100 Livorno, Italy\\
\email{pacini@di.unipi.it}
\and
 MARIA I. SESSA\\
DMI Universita' di Salerno - via S. Allende, 84081 Baronissi (SA), Italy\\
\email{mis@unisa.it}}

\maketitle

\begin{abstract}
In \cite{6.FPS95} an extended form of resolution, called Reduced SLD
resolution (RSLD), is introduced. In essence, an RSLD derivation is an SLD
derivation such that redundancy elimination from resolvents is performed
after each rewriting step. It is intuitive that redundancy elimination may
have positive effects on derivation process. However, undesiderable effects
are also possible. In particular, as shown in this paper, program
termination as well as completeness of loop checking mechanisms via a given
selection rule may be lost. The study of such effects has led us to an
analysis of selection rule basic concepts, so that we have found convenient
to move the attention from rules of atom selection to rules of atom
scheduling. A priority mechanism for atom scheduling is built, where a
priority is assigned to each atom in a resolvent, and primary importance is
given to the event of arrival of new atoms from the body of the applied
clause at rewriting time. This new computational model proves able to
address the study of redundancy elimination effects, giving at the same time
interesting insights into general properties of selection rules. As a matter
of fact, a class of scheduling rules, namely the specialisation independent
ones, is defined in the paper by using not trivial semantic arguments. As a
quite surprising result, specialisation independent scheduling rules turn
out to coincide with a class of rules which have an immediate structural
characterisation (named stack-queue rules). Then we prove that such
scheduling rules are tolerant to redundancy elimination, in the sense that
neither program termination nor completeness of equality loop check is lost
passing from SLD to RSLD.
\end{abstract}


\section{Introduction}

Several different approaches have been considered so far to enrich the SLD
resolution in order to improve the performance of top-down interpreters. The
usual objective is to reduce the search space without loss of results of the
refutation process, possibly obtaining a finite search space. Among the
proposed methods, the loop check mechanisms \cite{2.ABK89}, \cite{4.BAK91}, 
\cite{11.SGG}, \cite{13.vG87},\ and the tabulation technique \cite{BD98}, 
\cite{D87}, \cite{RRSSW99}, \cite{TS86}, \cite{14.V89}, aim to eliminate
redundant computations and to enforce the termination of a query over a
logic program.

Loop check mechanisms provide the interpreter with the capability of pruning
certain nodes of the SLD tree. The pruning is based on excluding some kinds
of structural repetitions for the goals in a derivation path. When suitable
structure repetitions are found, further rewritings of the current node are
ignored, because any solution possibly existing in the cut sub-tree is also
present in other parts of the SLD tree. Different forms of loop checks are
proposed in the literature. In particular, Bol et al. have defined several 
\emph{simple} loop checks, i.e. loop checks whose pruning mechanisms do not
depend on the considered logic program, and have analysed them against the
basic property of soundness and completeness \cite{4.BAK91}. The
completeness property concerns with the capability of pruning every infinite
derivation. In contrast, soundness concerns with the preservation of the
computed answer substitutions.

The main idea of tabulation originates from functional programming and
consists in building a table during the search of answers in an SLD tree.
The table contains entries for atoms with the corresponding answers so far
computed. These answers are to be used later, when instances of such atoms
should be recomputed. Such instantiated occurrences are named \emph{%
non-admissible atoms }(or \emph{consumer}). In essence, non-admissible atoms
are not resolved against clauses but against answers computed in other parts
of the SLD tree. The re-using approach exploited by the tabulation technique
was already mentioned by Kowalski \cite{K79} and has been proposed several
times under different names, such as memo-isation \cite{D87}, and
AL-technique \cite{14.V89}.

The conceptual differences between loop checks and tabulation are reflected
in several interesting aspects. In particular, tabulation requires a local
selection rule to guarantee the answer preservation, while no missing of
solution is possible with (sound) loop checks independently of the used
selection rule. On the other hand, the tabulation technique ensures
termination for any function-free program and for any program with a finite
Herbrand model, while the completeness of loop checks takes place for
specific classes of programs possibly with respect to given selection rules 
\cite{5.B92}, \cite{4.BAK91}, \cite{12.PS}. Finally, loop checks exploit no
auxiliary data structure and the pruning decision usually depends on the
current derivation only, while tabulation needs a table to store the answers
of atoms solved in the previously traversed portion of the tree.

Proposals can be also found in literature for a synergistic use of different
techniques aiming to optimise the query evaluation procedure. In particular,
in \cite{14.V89} a loop checking mechanism is combined with the tabulation
technique in order to eliminate some redundant parts of the search space. In 
\cite{6.FPS95} the simple loop check mechanisms proposed in \cite{4.BAK91}
are combined with another form of redundancy elimination which is named
(goal) \emph{reduction}. Goal reduction is conceptually analogous to the 
\emph{condensing} technique proposed by Joyner for the proof of the
unsatisfiability of first-order formulas \cite{8.J76}. In both cases
redundant atoms are eliminated from resolvents, in order to avoid useless
computations and to contain the size of the resolvents at the same time. The
main idea of reduction originates from the observation that if there exists
a refutation for an atom, then a refutation exists also for any more general
version of that atom. In this sense, such more general versions can be seen
as potentially redundant and we can imagine to remove them from the
resolvent, though suitable cares are to be taken as discussed in \cite
{6.FPS95}. By goal reduction, a generalised form of SLD resolution (named
RSLD) can be obtained, where a reduction of the resolvent is performed after
each rewriting step.

Goal reduction technique has a modus operandi 
which shows evident affinity with the one of loop checking mechanisms.
Indeed, with reduction
redundant atoms are definitively ignored, as it is done with loop checks for
pruned nodes. This is not the case with tabulation, in the sense that
non-admissible atoms, which are indeed solved against previously tabulated
answers, are not redundant. Such different philosophy between tabulation and
RSLD is highlighted also by the fact that the reduction technique eliminates
atoms in their \emph{more general version}, while non-admissible atoms are 
\emph{instances} of previously solved goals. It is evident that RSLD does
not need any auxiliary data structure because it considers only the current
goal (not even the current derivation path). The soundness of RSLD is shown
in \cite{6.FPS95} independently of the used selection rule. This means that
RSLD does not require particular selection rules in order to ensure answer
preservation.

It is intuitive that redundancy elimination may have positive effects on
derivation process. In \cite{6.FPS95}, advantageous combinations are shown
with respect to loop checking mechanisms. In particular, it is proven that a
well known simple loop check mechanism, namely Equality Variant check of
Resultant as Lists ($EVR_{L}$), becomes complete for several classes of
programs, provided that RSLD is exploited instead of usual SLD. The specific
reason is that the length of resolvents can be maintained within the limit
of a finite value through systematic elimination of redundant atoms. In
essence, there is clear evidence that the strength of equality loop checks
can augment if RSLD resolution is used.

However, even though not completely intuitive, redundancy elimination can
produce undesirable effects, too. In fact, as exemplified later, problems
can arise with program termination, as well as with the completeness of loop
checking mechanisms. The rationale behind this is that redundancy
elimination can affect the actual sequence of atom rewriting with respect to
given selection rules. This can (infinitely) delay the selection of failing
atoms, so that termination is missed. On the other hand, the structure of
the obtained resolvents can be altered by redundancy elimination, so that
loop checks may become unable to detect infinite derivations.

As shown in this paper, missing termination and loop detection depends
critically on the used selection rule. We say in the sequel that a selection
rule is \emph{redundancy elimination tolerant} if no loss in termination
and/or loop detection comes out, passing from SLD to RSLD.

In Section \ref
{s1.goalRedu}, we prove that termination and $EVR_{L}$ completeness are
preserved if they hold in SLD with respect to all possible selection rules.
Then, a more accurate analysis of redundancy elimination tolerance is
performed. To this aim, a careful reconsideration of selection rule basic
concepts will be required, so that we will be led to a reformulation of
selection rule ideas in terms of their operational counterparts, namely 
\emph{scheduling mechanisms}, so that we will prefer to talk of tolerant
scheduling rules. As a matter of fact, in Section \ref{s2.prioSche} we
provide a highly expressive execution model based on priority mechanism for
atom selection. A priority is assigned to each atom in a resolvent, and
primary importance is given to the event of arrival of new atoms from the
body of the applied clause at rewriting time. Indeed, new atoms can be
freely positioned with respect to the old ones in the resolvent, through the
assignment of priority values according to a given scheduling rule. Then, at
any derivation step, the atom with optimum priority is simply selected.

This new computational model proves able to address the study of redundancy
elimination effects, giving at the same time interesting insights into
general properties of selection rules. As a matter of fact, in Section \ref
{s3.specFree} a class of scheduling rules, namely the \emph{specialisation
independent} ones, is defined by using not trivial semantic arguments.
Several properties of specialisation independent scheduling rules are also
proven. As a quite surprising result, in Section \ref{s4.stacQueu} we show
that specialisation independent scheduling rules coincide with \emph{%
stack-queue rules}, which have an immediate structural characterisation.
Indeed, the stack-queue scheduling technique is simply defined so that, in
order to obtain the new resolvent at rewriting time, part of new atoms are
stacked at the beginning of the old resolvent while the remaining ones are
queued. Then in Section \ref{s5.reduElim} we prove that such scheduling
rules are tolerant to redundancy elimination, in the sense that neither
program termination nor completeness of equality loop check is lost passing
from SLD to RSLD. The proof is largely based on properties which we have
established for specialisation independent (and stack-queue) scheduling
rules.

\section{\label{s1.goalRedu}Goal reduction, program termination and\newline
$EVR_{L}$ completeness}

Throughout the paper we assume familiarity with the basic concepts of Logic
Programming \cite{1.A90}, \cite{A98}, \cite{10.L87}.

Here, only some notations are given about SLD derivation procedure, 
which can be described as follows. 
Let $G=a_{1},a_{2},...a_{k}$ be a \emph{goal}, constituted by a
conjunction of $k$ atoms, and $c=(ht\longleftarrow B)$ a \emph{clause}, where%
$\;ht\;$is an atom and$\;B\;$is a goal. The goal $G^{\prime }$ is a \emph{%
resolvent} of$\;G\;$and$\;c\;$by a renaming $\xi $ and a substitution $%
\theta $, if an atom $a_{i}$ exists, with $1\preceq i\preceq k$, such that $%
G^{\prime }=(a_{1},...a_{i-1},B\xi ,a_{i+1},...a_{k})\theta $, where $\theta 
$ is an idempotent and relevant mgu of $(ht)\xi $ and $a_{i}$. In the
sequel, given an expression $E$, the notation $var(E)$ will indicate the set
of variables in $E$. Moreover, we will denote by $(G\overset{c\xi ,\theta }{%
\longrightarrow }G^{\prime })$ the fact that $G^{\prime }$ is a resolvent of$%
\;G\;$and$\;c\;$by $\xi $ and $\theta $. Given an \emph{initial goal} $G_{o}$
and a logic program P, an \emph{SLD derivation of }$G_{o}$\emph{\ in P} is a
possibly infinite sequence of the type:\smallskip

$G_{o}\overset{c_{o}\xi _{o},\theta _{o}}{\longrightarrow }G_{1}\;...\;G_{j}%
\overset{c_{j}\xi _{j},\theta _{j}}{\longrightarrow }G_{j+1}...$\smallskip

\noindent such that, for any $j\succeq 0$, each clause $c_{j}$ belongs to P
and each $c_{j}\xi _{j}$ is \emph{standardised apart}, i.e.\smallskip

$var(c_{j}\xi _{j})\cap (var(G_{o})\cup var(c_{o}\xi _{o})\cup ...\cup
var(c_{j-1}\xi _{j-1}))=\varnothing .$\smallskip

A \emph{selection rule} is a function which chooses the atom to be rewritten
in the last resolvent of any finite SLD derivation. Given a selection rule$%
\;S,$ an SLD derivation is \emph{via}$\;S\;$if all the selections of atoms
are performed in agreement with$\;S.$ An SLD \emph{refutation} is a finite
SLD derivation such that the last resolvent is empty.\medskip

Now we can introduce the definitions of \emph{goal reduction} and RSLD
derivation. The reduction technique aims to eliminate redundant atoms from
the resolvents in order to contain their size. Analogous issue was already
been faced for the proof of the unsatisfiability of first-order formulas.
Indeed Joyner \cite{8.J76} noted that the increase in size of resolvents is
a factor which prevents resolution strategies being decision procedures for
solvable classes of first-order formulas (i.e. classes of formulas for which
the question of satisfiability or unsatisfiability can be effectively
decided). To limit the growth of the number of literals, Joyner introduced a
technique for simplifying resolvents, called \emph{condensing}. The
condensation of a clause is defined as the smallest subset of the clauses
which is also an instance of it. In other words, the condensation of a
clause can be obtained by applying a substitution $\alpha $ and eliminating
all the atom repetitions.

With reference to SLD derivations, the most evident form of redundancy
corresponds to multiple occurrences of the same atom in a resolvent. It is
obvious that this kind of atom repetition is essentially redundant. However,
this is not the only possible case of redundancy. Indeed, the reduction
technique, which is introduced in \cite{6.FPS95} as a variant of Joyner's
condensing technique, is able to perform quite general actions of redundancy
elimination from resolvents while preserving the soundness and the
completeness of RSLD resolution. By condensation, Joyner obtains a complete
and sound resolution procedures, which work as decision procedures for
several solvable classes of first order formulas \cite{8.J76}. By reduction,
the well known sound $EVR_{L}$ loop check becomes complete for several
classes of logic programs \cite{6.FPS95}.

Intuitively, the basic idea of goal reduction technique can be explained as
follows. Suppose having to refute a resolvent which contains $p(x)$ and $%
p(a) $, where $x$ is a variable and $a$ is a constant. Obviously, any
refutation for $p(a)$ implies a refutation for the atom $p(x)$, as $p(x)$ is
more general than $p(a)$. In this sense, the atom $p(x)$ may appear as a
redundant one. Actually, in order to ensure the soundness of the derivation
process, the elimination of redundant atoms (such as $p(x)$ above) is
conditioned in two aspects which can be sketched through the following
simple examples:

\begin{enumerate}
\item[a)]  Consider a resolvent like $p(x),q(x),p(a)$. In this case, the
atom $p(x)$ cannot be eliminated, because the connection between the atoms $%
p(x)$ and $q(x)$, by the variable $x$, is lost.\medskip

\item[b)]  Suppose that $x$ is a variable in the initial goal of a
derivation, and the actual resolvent is $p(x),p(a)$. In this case $p(x)$
cannot be dropped, because possible instantiations of $x$ in computed
answers could be lost. So we would obtain computed answers which are too
general with respect the correct answers, thus missing soundness.\medskip
\end{enumerate}

Now we present a formal definition of goal reduction which takes into
account the observations a) and b) and follows the line of Definition 2.1
presented in \cite{6.FPS95}. We will denote by $\subseteq _{L}$ the
inclusion relation between goals, and $G-N$ will indicate the goal obtained
from $G$ by eliminating the atoms which are present in $N$. In both cases
the goals are regarded as lists.\medskip

\begin{definition}[Reduced goal]
\label{d1.1}
Let $X$ be a set of variables, $\tau $ a
substitution and$\;G\;$a goal. A goal$\;N\;$is a \emph{reduced goal} of$%
\;G\; $by $\tau $ up to $X$, denoted by $G>>^{\tau }N$, if the following
conditions hold:\smallskip

i) $N\subseteq _{L}G,$\smallskip

ii) $\forall b\in (G-N),\;\;b\tau \in N,$\smallskip

iii) $\forall x\in (var(N)\cup X)\;\;$it is $x\tau =x.$\medskip
\end{definition}

In agreement with the above definition, a part $(G-N)$ of atoms of$\;G\;$can
be eliminated if a substitution $\tau $ exists such that $b\tau \in N$, for
any atom $b\in (G-N)$, provided that $\tau $ does not affect neither the
variables in$\;N\;$nor those in $X$. The imposition that $\tau $ does not
affect the variables in$\;N\;$prevents the kind of difficulties which are
exemplified in a).

\begin{example}
\label{e1.1}
\noindent Let:\smallskip

$G=p(z,v),q(w),p(w,v),p(w,x),p(w,y),q(v),q(y),$\smallskip

$X=\{x,w\}.$\smallskip

\noindent The following goal$\;N\;$is a reduced goal of$\;G\;$by $\tau
=\{z/w,y/v\}$ up to $X$:\smallskip

$N=q(w),p(w,v),p(w,x),q(v).$ \ \ $\square$\medskip
\end{example}

Performing reductions in the resolvents of an SLD derivation corresponds to
an actual extension of the SLD resolution process. Then, a generalised
version of SLD resolution can be introduced, i.e. the \emph{Reduced SLD
resolution} (RSLD in the sequel), where at any resolution step a reduction
of the resolvent is allowed. The following is the formal definition of RSLD
derivations.\medskip

\begin{definition}[Reduced SLD derivation]
\label{d1.2}
Let $P$ be a program and $G_{o}$
a goal. A \emph{Reduced SLD derivation} of $G_{o}$ in $P$ (RSLD in the
following) is a possibly infinite sequence of the form:\smallskip

$G_{o}>>^{\alpha _{o}}N_{o}\overset{c_{o}\xi _{o},\theta _{o}}{%
\longrightarrow }G_{1}\;...\;G_{h}>>^{\alpha _{h}}N_{h}\overset{c_{h}\xi
_{h},\theta _{h}}{\longrightarrow }G_{h+1}>>^{\alpha _{h+1}}N_{h+1}...$
\smallskip

\noindent where, for any $j\succeq 0,$\smallskip

i) $c_{j}$ is a clause in $P$,\smallskip

ii) $var(c_{j}\xi _{j})\cap (var(G_{o})\cup var(c_{o}\xi _{o})\cup ...\cup
var(c_{j-1}\xi _{j-1}))=\varnothing ,$\smallskip

iii) $G_{j}>>^{\alpha _{j}}N_{j}$ up to $var(G_{o}\theta _{o}...\theta
_{j-1}).$\medskip
\end{definition}

It is evident that an SLD derivation is a particular case of RSLD derivation
where $G_{j}=N_{j}$, for any $j$. Each $N_{j}$ is called a \emph{reduced
resolvent}. Condition ii) above is the usual standardisation apart
requirement. Condition iii) prevents the kind of difficulties which are
exemplified in b), guaranteeing the soundness of the mechanism. The
soundness and completeness of RSLD resolution are proven in Theorems 2.1 and
2.2 of \cite{6.FPS95}.

\subsection{\label{subs1.1}Program termination}

The completeness of RSLD resolution ensures that missing computed answers is
impossible when we pass from SLD to RSLD. This is not the case with
termination, as shown by the following Example \ref{e1.1.1}. In the example
a selection rule$\;S\;$and a program P are given, such that any SLD
derivation of P via$\;S\;$terminates independently of the initial goal.
However, we show that termination is lost, if reduction of resolvents is
performed.

\begin{example}
\label{e1.1.1}
\noindent Let us consider a selection rule$\;S\;$such that, given a goal $G$, 
the first atom is chosen for rewriting if the length of$\;G\;$is odd, and
the last atom is chosen otherwise. Let us consider the logic program P
consisting of the following clause:\smallskip

$c=p(x,y)\longleftarrow q,p(x,z_{1}),p(z_{1},z_{2}),p(z_{2},y).$\smallskip

\noindent It can be easily seen that all SLD derivations in P via$\;S\;$%
terminate, independently of the initial goal. Indeed, suppose that the
initial goal has an odd number of atoms. It is evident that either the
derivation via$\;S\;$fails immediately or the initial goal has the form ``$%
p(..),Y$'', so that the first step of the derivation produces a resolvent of
an even length as follows:\smallskip

$p(..),Y\overset{c}{\longrightarrow }q,p(..),p(..),p(..),Y.$\smallskip

\noindent Now, either the derivation fails immediately or $Y=Z,p(..)$, so
that a second derivation step is performed:\smallskip

$q,p(..),p(..),p(..),Z,p(..)\overset{c}{\longrightarrow }%
q,p(..),p(..),p(..),Z,q,p(..),p(..),p(..)$,\smallskip

\noindent and the process fails anyway, since the last resolvent has an odd
length. Then, suppose on the contrary that the initial goal has an even
number of atoms. Either the derivation fails immediately or the initial goal
has the form ``$T,p(..)$''. In the second case, the first derivation step
gives place to a resolvent with an odd length, so that the derivation fails.
Now, let us verify that termination can be lost if reduction of resolvents
is performed. Indeed, let us consider the RSLD derivation of the
goal $(q,p(x,x))$ in P via $S$ given in Figure \ref{f1}. 
It is evident that the number of atoms is
even in any reduced resolvent. Thus, the last atom is always selected and
the derivation is infinite.~$\square$
\begin{figure}[h]
\raggedright \qquad \qquad \underline{Resolvents}\qquad  
\qquad \underline{Reduced Resolvents}\smallskip \\

\qquad \qquad  $q,p(x,x)$\smallskip \\

\qquad \qquad $\qquad \qquad >> q,p(x,x)$\qquad $\overset{c}
{\longrightarrow }$\smallskip \\

\qquad \qquad $q,q,p(x,z_{1}),p(z_{1},z_{2}),p(z_{2},x)$
\smallskip \\

\qquad \qquad $\qquad \qquad >> q,p(x,z_{1}),p(z_{1},
z_{2}),p(z_{2},x)$\qquad $\overset{c}
{\longrightarrow }$\smallskip \\

\qquad \qquad $q,p(x,z_{1}),p(z_{1},z_{2}),q,
p(z_{2},z_{3}),p(z_{3},z_{4}),
p(z_{4},x)$\smallskip \\

\qquad \qquad $\qquad \qquad
>> q,p(x,z_{1}),p(z_{1},z_{2}),p(z_{2},z_{3}),
p(z_{3},z_{4}),p(z_{4},x)$
\qquad $\overset{c}{\longrightarrow }$\smallskip \\

\qquad \qquad ........................
.....................
\caption{} \label{f1}
\end{figure}
\end{example}

As shown by the example in Figure \ref{f1}, termination with respect to a given selection
rule can be missed, if we pass from SLD to RSLD resolution. On the contrary,
we show in this section (Theorem \ref{t1.1.1}) that termination is
preserved, when any SLD derivation of$\;G\;$in P is finite independently of
the used selection rule. Theorem \ref{t1.1.1} will be proven as an immediate
consequence of the following Lemma \ref{L1.1.1}

\begin{lemma}
\label{L1.1.1} Let P be a program and $G_{o}$ a goal. For any possibly
infinite RSLD derivation $D$ of $G_{o}$ in P, an SLD derivation $D^{\prime }$
of $G_{o}$ in P exists, such that every reduced resolvent of $D$ is included
in the corresponding resolvent of $D^{\prime }$ up to renamings.
\end{lemma}

\begin{proof*}
Consider a possibly infinite RSLD derivation $D$
of $G_{o}$ in P\smallskip

$D=(G_{o}>>^{\alpha _{o}}N_{o}\overset{c_{o}\xi _{o},\theta _{o}}{%
\longrightarrow }G_{1}...$\smallskip

$\qquad \qquad ...G_{h}>>^{\alpha _{h}}N_{h}\overset{c_{h}\xi _{h},\theta
_{h}}{\longrightarrow }G_{h+1}>>^{\alpha _{h+1}}N_{h+1}...)$\hfill (1)\qquad
\smallskip

\noindent Intuitively, the SLD derivation $D^{\prime }$ is obtained by
choosing, step by step, the same clause and the same atom as in $D$. This
way, redundant atoms are not eliminated from resolvents of $D^{\prime }$,
but they have no real influence on the derivation process. More formally,
suppose that an SLD derivation of $G_{o}$ in P is already constructed
like\smallskip

$G_{o}\overset{c_{o}\xi _{o}^{\prime },\theta _{o}^{\prime }}{%
\longrightarrow }G_{1}^{\prime }...\longrightarrow G_{i}^{\prime },\hfill $%
(2)\qquad \smallskip

\noindent such that, for any $0\preceq j\preceq i$, a renaming $\tau _{j}$
exists with $N_{j}\tau _{j}\subseteq _{L}G_{j}^{\prime }.$ It is easy to
show that derivation (2) can be extended of one step in agreement with the
lemma. Let $a$ be the atom which is rewritten in the step $N_{i}\overset{%
c_{i}\xi _{i},\theta _{i}}{\longrightarrow }G_{i+1}$ of derivation (1). It
is evident that the clause $c_{i}$ is applicable to the atom $a\tau _{i}\in
N_{i}\tau _{i}\subseteq _{L}G_{i}^{\prime },$ so that we have an SLD
derivation step of the form:\smallskip

$G_{i}^{\prime }\overset{c_{i}\xi _{i}^{\prime },\theta _{i}^{\prime }}{%
\longrightarrow }G_{i+1}^{\prime }.\hfill $(3)\qquad \smallskip

\noindent Now let $F$ denote the sublist of atoms in $G_{i+1}^{\prime }$
which derives from $N_{i}\tau _{i}$. It is obvious that the subgoal $%
(G_{i}^{\prime }-N_{i}\tau _{i})$ has no active role in derivation step (3).
So, we have that $F$ is a variant of $G_{i+1}$, i.e. a renaming $\tau _{i+1}$
exists with $F=G_{i+1}\tau _{i+1}$, which means that $G_{i+1}\tau
_{i+1}\subseteq _{L}G_{i+1}^{\prime }$. But, by definition of goal reduction
we have $N_{i+1}\subseteq _{L}G_{i+1}.$ As a consequence\smallskip 

$N_{i+1}\tau_{i+1}\subseteq _{L}G_{i+1}\tau _{i+1}
\subseteq _{L}G_{i+1}^{\prime }.$ \ \ $\square$\medskip
\end{proof*}

\begin{theorem}
\label{t1.1.1}Let P be a program and$\;G\;$a goal. If every SLD derivation of%
$\;G\;$in P is finite independently of the used selection rule, then every
RSLD derivation of$\;G\;$in P is finite too.
\end{theorem}

\begin{proof}
Suppose that an infinite RSLD derivation of$\;G\;$
in P exists. By Lemma \ref{L1.1.1}, an infinite SLD derivation of$\;G\;$in P
also exists, which contradicts the hypothesis.
\end{proof}

\subsection{\label{subs1.2}$EVR_{L}$ loop check completeness}

The termination issue of a query to a logic program has attracted much
attention over the past few years, both in the logic programming field, and
in the deductive database field (see \cite{DD94} for a survey).

A well known approach to the termination problem of 
a query in a logic program consists
in modifying the computation mechanism by adding a capability of pruning,
i.e. at some point the interpreter is forced to stop its search through a
certain part of the SLD tree \cite{2.ABK89}, \ \cite{5.B92}, \ \cite{4.BAK91}, 
\ \cite{12.PS}, \ \cite{11.SGG}, \ \cite{13.vG87}. These mechanisms are called 
\emph{loop checks}, as they are based on discovering some kinds of
repetitions in derivation paths. The purpose of a loop check is to reduce
the search space for top-down interpreters in order to prune infinite
derivations, without loss of results of the refutation process. Thus, two
basic properties are considered for loop checks. The \emph{completeness}
property of a loop check concerns the capability of pruning every infinite
derivation. In contrast, the \emph{soundness} property has to do with the
preservation of computed answer substitutions.

Different forms of loop checking are considered in literature. A systematic
analysis of loop checking for SLD resolution is given in \cite{4.BAK91}. 
\emph{Simple loop checks} have deserved special interest, because the
decision of pruning does not depend on the logic program we are confronted
with. The more immediate form of simple and sound loop check is the so
called \emph{Equality Variant of Resultant} check, which requires the
detection of equal (up to renaming) resultants in the derivation. Such a
loop check is formulated with respect to RSLD derivations in the following
Definition \ref{d1.2.1} which recalls the essence of the analogous
Definition 3.19 in \cite{6.FPS95}. The notation $(F=_{L}G)$ is used, which
means that the goal $F$ is equal to $G$, where the goals are regarded as
lists.
\bigskip

\begin{definition}[Equality Variant Check for Resultants]
\label{d1.2.1} 
An RSLD
derivation\smallskip

$G_{o}>>^{\alpha _{o}}N_{o}\overset{c_{o}\xi _{o},\theta _{o}}{%
\longrightarrow }G_{1}...G_{h-1}>>^{\alpha _{h-1}}N_{h-1}\overset{c_{h-1}\xi
_{h-1},\theta _{h-1}}{\longrightarrow }G_{h}>>^{\alpha _{h}}N_{h}...$%
\smallskip

\noindent is \emph{pruned} by \emph{Equality Variant of Resultant as Lists}
loop check ($EVR_{L}$ in the following), if for some $i$ and $j$, with $%
0\preceq i<j$, a renaming $\tau $ exists such that:\smallskip

i) $G_{o}\theta _{o}...\theta _{j-1}=G_{o}\theta _{o}...\theta _{i-1}\tau ,$%
\smallskip

ii) $N_{j}=_{L}N_{i}\tau .$\medskip
\end{definition}

Given an RSLD tree T, the application of $EVR_{L}$ yields a prefix Tp of T
which is obtained in this way. The descendants of a node are thrown away iff
the derivation associated with the path from the root to the node is pruned.

Any couple $Rs_{h}=[N_{h},G_{o}\theta _{o}...\theta _{h-1}]$ is a \emph{%
reduced resultant}. Given two resultants $Rs_{j}=[N_{j},G_{o}\theta
_{o}...\theta _{j-1}]$ and $Rs_{i}=[N_{i},G_{o}\theta _{o}...\theta _{i-1}],$
for which requirements i) and ii) of Definition \ref{d1.2.1} hold, we will
write $Rs_{i}\cong _{L}Rs_{j}.$ In other words, Definition \ref{d1.2.1}
expresses that $EVR_{L}$ check is based on detecting that a resultant is
obtained which is related by $\cong _{L}$ to a preceding one in the same
derivation. It is worth noting that the relationship $\cong _{L}$ is an
equivalence relationship. It is evident that, if reduction of resolvents is
always ineffective (i.e. $G_{j}=N_{j}$, for any $j$), the usual $EVR_{L}$
loop check for SLD derivations is found again. It is well known that $%
EVR_{L} $ is a sound loop check in the case of SLD resolution. The soundness
of $EVR_{L}$ is extended to the more general case of RSLD by Theorem 4.1 of 
\cite{6.FPS95}.

Let us observe that if we do not consider condition i) in Definition \ref
{d1.2.1} we obtain the $EVG_{L}$ loop check which is based on detecting that
a resolvent is obtained which is a variant of a preceding one in the same
derivation. It is worth noting that $EVG_{L}$ is a \emph{weakly sound} loop
check, in sense that it preserves at least a successful, but it does not
ensure the preservation of the computed answer substitutions \cite{4.BAK91}.

The completeness of a loop check is usually referred to given selection
rules and classes of programs. A loop check is complete for a program P with
respect to a selection rule$\;S\;$if all infinite derivations of P via$\;S\;$%
are pruned. A loop check is complete for a class C of programs, if it is
complete for every program in C. Several classes of logic programs are
characterised in literature for which complete loop checks can be found.
Actually, most of them are classes of function free programs, i.e. programs
whose clauses contain no function symbol \cite{5.B92}, \cite{4.BAK91}, \cite
{6.FPS95}, \cite{12.PS}. In the following of this section, and later in
Section \ref{s5.reduElim}, we consider the problem of preserving the
completeness of $EVR_{L}$ check, passing from SLD to RSLD resolution, in the
case of function free programs.

Let us first show how the completeness of equality loop checks, with respect
to a given selection rule, can be lost passing from SLD to RSLD. Indeed, it
is sufficient reconsider Example \ref{e1.1.1}. In that case $EVR_{L}$ loop
check is obviously complete, since no infinite SLD derivation exists. On the
other hand, it is obvious that $EVR_{L}$ loop check cannot prune the
infinite RSLD derivation developed in the same example, because the length
of resolvents increases at each derivation step. Actually, it is immediate
to verify that the infinite derivation in Example \ref{e1.1.1} cannot even
be pruned by using more complex and powerful checks (like $SIR_{M}$) which
are based on \emph{subsumption} relationships between resultants \cite
{4.BAK91}.

Now we prove that $EVR_{L}$ loop check completeness is preserved for
function free programs, in the case that $EVR_{L}$ is complete with respect
to all selection rules. Precisely, Theorem \ref{t1.2.1} states that, if $%
EVR_{L}$ prunes every infinite SLD derivation of a goal$\;G\;$in a function
free program P, then $EVR_{L}$ prunes also every infinite RSLD derivation of$%
\;G\;$in P. In order to show this result, let us provide a condition which
holds whenever $EVR_{L}$ prunes every infinite derivation of$\;G\;$in P.
Lemma \ref{L1.2.1} states that, if $EVR_{L}$ check prunes all infinite
derivations of$\;G\;$in P, then the length of resolvents in all possible
derivations is limited. In the proof of Lemma \ref{L1.2.1} we exploit the
notion of S-tree \cite{3.AP93}. Given a program P and a goal $G$, an S-\emph{%
tree} of$\;G\;$in P is a tree where the descendants of a goal are its
resolvents with respect to all selection rules and all input clauses. In
other words, an S-tree groups all SLD derivations of$\;G\;$in P. The
notation $\#R$ represents the number of atoms in the goal $R$.

\begin{lemma}
\label{L1.2.1}Let P be a program and$\;G\;$a goal. Suppose that all infinite
SLD derivations of$\;G\;$in P are pruned by $EVR_{L}$. Then, a finite bound $%
l$ exists such that, for each resolvent$\;R\;$in any SLD derivation of$\;G\;$%
in P, it is $\#R\preceq l$.
\end{lemma}

\begin{proof}
Let T be an S-tree of$\;G\;$in P. Given a node$\;n$%
\ in T, let $Dr(n)$ denote the derivation associated to the path from the
root of T to $n$, and $R(n)$ the final resolvent of $Dr(n)$. Then, let Tp be
the prefix of T which is obtained by applying the $EVR_{L}$ check to T, i.e.
the prefix where the descendants of any node$\;n$\ of T are thrown away if
and only the derivation $Dr(n)$ is not pruned by $EVR_{L}$. By hypothesis,
all infinite SLD derivations of$\;G\;$in P are pruned by $EVR_{L}$, which
means that Tp has no infinite path. As a consequence, since T is a finitely
branching tree, by Konig's lemma (see Theorem K, in \cite{9.K97}) the prefix
Tp is finite. Now, let $d$ be the depth of Tp, and $l$ the maximum of the
set $\{\#R(n)|\;n$\ is a node in Tp$\}$. We prove that:\smallskip

$\#R(n)\preceq l,$ for any node$\;n$\ in T.\smallskip

\noindent The proof is by induction on the value of $depth(n)$. For $%
depth(n)\preceq d$ the thesis is trivial. Then consider an integer $h>d$,
and suppose that $\#R(n^{\prime })\preceq l$, for any node $n^{\prime }$
with $depth(n^{\prime })<h.$ Given a node$\;n$\ of T such that $depth(n)=h$,
we show that also $\#R(n)\preceq l$ holds. Since$\;n\notin $Tp, the
derivation $Dr(n)$ is pruned by $EVR_{L}$, so that two nodes $n_{1}$ and $%
n_{2}$ exist in the path from the root of T to$\;n$\ with:\smallskip

- $depth(n_{1})<depth(n_{2}),$\hfill (1)\qquad \smallskip

- $R(n_{2})$ is a variant of $R(n_{1})$.\hfill (2)\qquad \smallskip

\noindent Now, consider the sequence of clauses which has determined the
path from $n_{2}$ to$\;n$\ in T. Since T contains all SLD derivations of$%
\;G\;$in P, the same derivation steps can be repeated in T starting from $%
n_{1}$. As a consequence, by (1) and (2), a path from $n_{1}$ to a node $%
n^{\prime }$ exists such that:\smallskip

- $depth(n^{\prime })=depth(n)-(depth(n_{2})-depth(n_{1}))<depth(n)=h$%
,\smallskip

- $R(n^{\prime })$ is a variant of $R(n)$.\smallskip

\noindent By inductive hypothesis it is $\;\#R(n^{\prime })\preceq l$. \ But 
$R(n^{\prime })$ is a variant of $R(n)$, so that$\;\;\#R(n)=\#R(n^{\prime
})\preceq l$. 

\noindent In conclusion, the thesis holds for every node$\;n$\ in T.\medskip
\end{proof}

\begin{theorem}
\label{t1.2.1}Let P be a function free program and$\;G\;$a goal. If $EVR_{L}$
prunes every infinite SLD derivation of$\;G\;$in P independently of the used
selection rule, then $EVR_{L}$ prunes every infinite RSLD derivation 
of $G$ in P.
\end{theorem}

\begin{proof}
Let $D$ be an infinite RSLD derivation of$\;G\;$in
P. By Lemma \ref{L1.1.1}, an SLD derivation $D^{\prime }$ of$\;G\;$in P also
exists such that every reduced resolvent of $D$ is included in a resolvent
of $D^{\prime }$ (up to renamings). Since $EVR_{L}$ prunes every infinite
SLD derivation of$\;G\;$in P, by Lemma \ref{L1.2.1} the length of resolvents
of $D^{\prime }$ is limited. Then, the length of reduced resolvents and
resultants of $D$ is also limited. Now, since the language of P is function
free and has finite many predicate symbols and constants, the relationship
denoted by $\cong _{L}$ has only finitely many equivalence classes on
resultants of $D$. As a consequence, for some $0\preceq i<k$ we have that
the $k^{th}$ and the $i^{th}$ resultants of $D$ are in $\cong _{L}$
relationship. This implies that $D$ is pruned by $EVR_{L}$.\medskip
\end{proof}

In this section, redundancy elimination tolerance has been proven on the
basis of a rather strong hypothesis, i.e. termination and completeness of
loop checking for all possible selection rules. In Section \ref{s2.prioSche}
we will introduce a new computational model which will allow us to
characterise a class of selection rules which are shown to be redundancy
elimination tolerant. As a matter of fact, in Section \ref{s5.reduElim} we
will prove that program termination and $EVR_{L}$ loop check completeness
are maintained for that class of rules, passing from SLD to RSLD.

\section{\label{s2.prioSche}Priority scheduling rules}

As shown in Section \ref{s1.goalRedu}, redundancy elimination can determine
missing termination and loop check detection. This fact depends critically
on the used selection rule, because redundancy elimination can affect the
actual sequence of atom rewriting. As a matter of fact, it is widely
acknowledged that the analysis of interdependence between derivation
processes and the used selection rules is a difficult task. In our study,
the necessary insights have been provided by a computation model which is
based on a novel mechanism of atom choice, which works in terms of \emph{%
scheduling rules} rather than in terms of conventional selection rules.
Through this new computational model, a class of scheduling rules is
identified in Section \ref{s3.specFree}, which is \emph{redundancy
elimination tolerant} in the sense that no loss in termination and/or loop
detection comes out, passing from SLD to RSLD.

We start the analysis with an observation about selection rules, as they are
normally conceived in literature and used in practice. In SLD derivations,
resolvents are usually regarded as lists, nevertheless selection rules are
given complete free choice ability of the atom to rewrite. In this sense,
two different philosophies are superimposed, because a scheduling (i.e. an
ordering) must coexist with an atom choice which can actually overcome the
scheduling. Now, in the case that resolvents are viewed as unstructured
multisets instead of lists, the obvious solution is that a free choice
ability is provided at rewriting time. But, if scheduling policies (i.e. an
ordering or a priority assignment) are exploited, it may appear natural that
priorities are obeyed at rewriting time, so that the atom with optimum
priority is always selected. Indeed, if a scheduling policy is used, the
moment of addition of new atoms in the resolvent may be recognised as the
really important event, when suitable priority values must be established
and assigned.

In the following of the paper we consider execution mechanisms for logic
programs which are based on priority scheduling policies. In particular we
characterise \emph{scheduling rules} informally as follows:

\begin{itemize}
\item  a priority value is assigned to each atom in the actual resolvent,

\item  assigned priorities are not modified in the following of the
derivation,

\item  the atom with optimum priority is always taken for rewriting.
\end{itemize}

In essence a scheduling rule is a rule that defines a priority values for
any new atom which enters the actual resolvent. It is crucial that atoms
from the body of the applied clause can be freely scheduled with respect to
the ones already present in the resolvent, which maintain their own priority
values. It is intuitive that this can be easily done if a set of ``dense''
priority values is adopted. Indeed, as formalised in Section \ref{subs2.1},
we will use rational numbers as priority values.

Now, in analogy with Lloyd's definition of selection rules \cite{10.L87}, we
consider the subclass of scheduling rules where the schedule of new atoms is
determined only by the last resolvent in the derivation, i.e. by the \emph{%
current state} of the computation. Such rules will be named state \emph{%
scheduling rules}. A state scheduling rule can be seen as a rule which, for
any resolvent $G$ and clause $c$ (that is applied to the optimum priority
atom), determines the schedule positions of the new atoms in the resolvent,
through the assignment of appropriate priority values.

In other words, a state scheduling rule determines new resolvents, starting
from the old ones and from applied clauses. The rewritten atom is
necessarily the one with the optimum priority value. It is evident that the
transformation from a resolvent to a new one, which is obtained by the
addition of new atoms from the applied clause, is nothing more than a step
of an SLD derivation. In this sense, we can say that a state scheduling rule
characterises a set of derivation steps. Indeed, as formalised in Section 
\ref{subs2.5}, a state scheduling rule can be straight conceived as a \emph{%
set of derivation steps}, that is: the set of derivation steps which are
allowed according to the scheduling rule itself. Formal definition of state
scheduling rules is provided in Section \ref{subs2.5}.

\subsection{\label{subs2.1}Atoms, goals and priorities}

In order to characterise \emph{state scheduling rules} in a formal way, we
introduce the notions of \emph{priority goal} and \emph{priority clause}. A
priority goal is a goal where each atom has an associated priority value.
Thus, a priority goal$\;G\;$can be thought as a set of couples, where any
couple is named \emph{priority atom}. In the following formal definition,
priority atom will be denoted by $a[p]$, where $a$ is an usual atom and $p$
is a rational number which establishes the priority of $a$ in $G$. The
symbol $\Longrightarrow $ will be frequently used in the rest of the paper
to denote logical implication.\smallskip

\begin{definition}
\label{d2.1.1}
\begin{enumerate}
\item[i)]  A \emph{priority goal}$\;G\;$(\emph{p-goal} in the sequel) is
defined by a set of \emph{priority atoms }(or simply \emph{p-atoms}) of the
form:\smallskip

$G=\{a_{1}[p_{1}],...a_{k}[p_{k}]\},$ with $\forall i,j:\;i\neq
j\Longrightarrow p_{i}\neq p_{j},$\smallskip

where each $a_{m}$ is an usual atom and each $p_{m}$ is a rational number, $%
1\preceq m\preceq k$.\medskip

\item[ii)]  A \emph{priority clause }(or simply a \emph{p-clause}) has the
form $c=ht\longleftarrow B$, where$\;ht\;$is an atom (without priority) and$%
\;B\;$is a priority goal.\medskip
\end{enumerate}
\end{definition}

In the sequel, priority clauses will be referred as clauses for the sake of
simplicity. Capital letters will be used in the following to represent
p-goals. In order to denote p-atoms, we will use notations like $a[p]$, as
well as simple small letters (as $a,b$, etc.) when explicit reference to
priority values is not important. As a slight abuse of notation, p-goals
made of only one p-atom $a$ will be often denoted by $a$. 
Given a p-goal $G$, the notations $\#G$ will indicate the number 
of p-atoms in $G$.

In the sequel, we will exploit very frequently a basic operation which
corresponds to the union of two p-goals with no common priority values. This
operation is denoted by ``+'' and is said p-goal \emph{merging}. During
merging operations, atoms retain their priority values. We introduce also
the idea of \emph{concatenation}, which is a particular case of merging.
Concatenations will be denoted by the symbol ``\TEXTsymbol{\vert}''
(vertical bar). The following are the formal definitions of merging and
concatenation. It is worth noting that both these operations are associative.

\begin{definition}
\label{d2.1.2} 
\begin{enumerate}
\item[i)]  A p-goal$\;M\;$is the \emph{merging} of$\;F\;$and$\;G\;$(denoted
by $M=F+G$) if$\;F\;$and$\;G\;$have no common priority values and 
$M=F\cup G$.\medskip

\item[ii)]  Given two p-goals$\;F\;$and $G$, we write$\;F\dashv G\;$to
denote that all priorities in$\;F\;$are less than any priority in $G.$ A
p-goal$\;N\;$is the \emph{concatenation} of$\;F\;$and$\;G\;$(denoted by $%
N=F|G$), if $N=F+G$ and$\;F\dashv G$.\medskip
\end{enumerate}
\end{definition}

The fact that equal priority values are not admitted in a p-goal has two
principal effects. The first one is that a complete ordering (i.e. a
scheduling) is imposed on the atoms of a p-goal. In particular we assume
that atoms with less priorities precede atoms with greater ones. The second
effect is that possible multiple occurrences of atoms are distinguished by
different priority values. On the basis of the above observations, the
following evident properties of concatenation can be stated.

\begin{property}
\label{py2.1.1}Given the p-goals $A_{1},A_{2},A_{3},B_{1},B_{2}$, 
and $B_{3}$, the following propositions hold:

\begin{enumerate}
\item[i)]  $A_{1}|A_{2}=B_{1}|B_{2},\;\#A_{1}=\#B_{1}$ or $%
\#A_{2}=\#B_{2}\Longrightarrow A_{1}=B_{1},\;A_{2}=B_{2}.$\medskip

\item[ii)]  $A_{1}|A_{2}|A_{3}=B_{1}|B_{2}|B_{3},\;A_{2}\neq \varnothing
,\;A_{2}=B_{2}\Longrightarrow A_{1}=B_{1},\;A_{3}=B_{3}.$
\end{enumerate}
\end{property}

\subsection{\label{subs2.2}Shifting and positioning}

Throughout the paper, we will exploit a basic operator for handling priority
values. It will be said (\emph{priority}) \emph{shifting}, and corresponds
to a modification of priority values which does not alter the scheduling of
the atoms in a p-goal. The following is the formal definition of shifting.
In the sequel, shiftings will be always denoted by underlined Greek letters.

\begin{definition}[shifting]
\label{d2.2.1}
A \emph{shifting} $\underline{\pi }$ is an
increasing one-to-one application of the type:\smallskip

$\underline{\pi }$ : Rational $\longrightarrow $ Rational.\smallskip

\noindent Given a \emph{shifting} $\underline{\pi }$, and two p-goals$\;G\;$%
and$\;F\;$such that:\smallskip

$G=\{a_{1}[p_{1}],...a_{k}[p_{k}]\}$ and $F=\{a_{1}[\underline{\pi }%
(p_{1})],...a_{k}[\underline{\pi }(p_{k})]\}$,\smallskip

\noindent we say that$\;F\;$is a \emph{shifting} of$\;G\;$and write$\;F=G%
\underline{\pi }$.\medskip
\end{definition}

It is evident that the composition of two shiftings is a shifting, too, as
well as the inverse of a shifting. Shifting operations enjoy the following
four basic properties. All properties are plain consequence of the
definition. The first two properties will be used very often in the sequel
without explicit reference.

\begin{property}
\label{py2.2.1}\smallskip

\noindent Ax-i) $\ \ \ \ \ (A_{1}+A_{2}+...+A_{k})\underline{\pi }=A_{1}%
\underline{\pi }+A_{2}\underline{\pi }+...\;A_{k}\underline{\pi },$\smallskip

\noindent Ax-ii) $\ \ \ \ (A_{1}|A_{2}|...\;A_{k})\underline{\pi }=A_{1}%
\underline{\pi }|A_{2}\underline{\pi }|...\;A_{k}\underline{\pi },$\smallskip

\noindent Ax-iii) $\ \ \ G=A_{1}\underline{\tau }_{1}|A_{2}\underline{\tau }%
_{2}|...\;A_{k}\underline{\tau }_{k},\;\;\;F=A_{1}\underline{\pi }_{1}|A_{2}%
\underline{\pi }_{2}|...\;A_{k}\underline{\pi }_{k}$\smallskip

$\qquad \qquad \qquad \Longrightarrow $\ $\exists $ $\underline{\sigma }$ \ 
\emph{such that} $\ \ F\underline{\sigma }=G$,\smallskip

\noindent Ax-iv) $\ \ \ (A_{1}+A_{2}+...\;A_{k})\underline{\pi }%
=(A_{1}+A_{2}+...\;A_{k})\underline{\tau }$\smallskip

$\qquad \qquad \qquad \Longrightarrow \;A_{1}\underline{\pi }=A_{1}%
\underline{\tau },\;\;A_{2}\underline{\pi }=A_{2}\underline{\tau }%
,\;\;...\;\;A_{k}\underline{\pi }=A_{k}\underline{\tau }.$\medskip
\end{property}

Finally let us consider a combination of shifting and merging which provides
the convenient tool to formalise our ideas about scheduling of atoms in
resolvents. As outlined in previous section, at any step of derivation,
atoms coming from the body of the applied clause are assigned new priority
values, while priorities of old atoms are left unchanged. This way, new
atoms are positioned (i.e. scheduled) with respect to the old ones. In
general, the \emph{positioning} of atoms from a p-goal $B$, with respect to
the atoms of another p-goal $F$, can be described through a composition of
shifting and merging. Indeed, consider an expression like $F+B\underline{\pi 
}$. The effect of the shifting $\underline{\pi }$ is twofold. First of all,
possible conflicts of priority values between$\;F\;$and$\;B\;$can be
removed, so that the merging $F+B\underline{\pi }$ is correctly performed.
At the same time, yet more important, $\underline{\pi }$ allows us to
establish the positions which atoms from$\;B\;$go to occupy. Since
priorities are represented by rational values, it is evident that all
possible allocations of atoms from $B$, with respect to those in $F$, can be
described through suitable choices of $\underline{\pi }$.

\subsection{\label{subs2.3}Priority SLD Derivations}

Now, we are ready to frame well known Logic Programming concepts, as the
ones of resolvent and SLD derivation, in terms of priority atoms, goals and
scheduling. We start with the following Definition \ref{d2.3.1}, which
formalises the idea of \emph{priority derivation step}. Given a p-goal $a|F$%
, in agreement with our concept of scheduling the atom $a$ with minimum
priority is always rewritten and atoms coming from the body of the applied
clause are positioned with respect to old ones to form the new resolvent.
The positioning is obtained through a combination of shifting and merging,
as discussed at the end of the previous Section \ref{subs2.2}. With
reference to Definition \ref{d2.3.1}, the body$\;B\;$of the applied clause
is first shifted by \underline{$\pi $} and then merged with $F$, i.e. with
the initial p-goal $a|F$ minus the rewritten atom.

\begin{definition}[priority derivation step]
\label{d2.3.1}
Consider a p-goal$\;G\;=a|F$
and a clause $c=(ht\longleftarrow B)$. Let:\smallskip

- $\xi $ be a renaming such that $var(G)\cap var(c\xi )=\varnothing $,
\smallskip

- $\theta $ be an idempotent and relevant mgu of $a$ and $(ht)\xi $,
\smallskip

-$\;\underline{\pi }$ be a shifting such that$\;F\;$and $B\underline{\pi }$
have no common priority value.\smallskip

\noindent We say that$\;R\;$is a \emph{resolvent} of$\;G\;$and$\;c\;$by $\xi 
$, $\theta $ and $\underline{\pi }$, if:\smallskip

$R=(F+B\xi \underline{\pi })\theta $.\smallskip

\noindent The transformation from $a|F$ to $(F+B\xi \underline{\pi })\theta $
will be called a \emph{priority derivation step}. It is denoted by:\smallskip

$a|F\overset{c}{\longrightarrow }(F+B\xi \underline{\pi })\theta $.\medskip
\end{definition}

The notation$\;G\;\overset{c\xi ,\theta }{\longrightarrow }R$ will be used
to represent a derivation step by $\theta $ and $\xi $, where the shifting 
\underline{$\pi $} is not pointed out. Analogously, we will write$\;G%
\overset{c\xi }{\longrightarrow }R$ to represent a derivation step by the
renaming $\xi $ without specifying the mgu $\theta $. By$\;G\overset{c}{%
\longrightarrow }R$ we denote a derivation step which generically produces$%
\;R\;$as a resolvent of$\;G\;$and $c$. Iterating the process of computing
resolvents, we obtain a priority SLD derivation, that is a sequence of
priority derivation steps as formalised by the following definition.

\begin{definition}[priority SLD derivation]
\label{d2.3.2}
Let P be a program and 
$G_{o} $ a p-goal. A \emph{priority SLD derivation} of $G_{o}$ in P is a
possibly infinite sequence of priority derivation steps\smallskip

$G_{o}\overset{c_{o}\xi _{o},\theta _{o}}{\longrightarrow }%
G_{1}\longrightarrow ...\;G_{k}\overset{c_{k}\xi _{k},\theta _{k}}{%
\longrightarrow }G_{k+1}\longrightarrow ...$\smallskip

\noindent where, for any $j\succeq 0$,\smallskip

i) $c_{j}$ is a clause in P,\smallskip

ii) $var(c_{j}\xi _{j})\cap (var(G_{o})\cup var(c_{o}\xi _{o})\cup ...\cup
var(c_{j-1}\xi _{j-1}))=\varnothing .$\medskip
\end{definition}

Given a finite priority SLD derivation (p-SLD \emph{derivation} in the
following) of the form:\smallskip

$G_{o}\overset{c_{o}\xi _{o},\theta _{o}}{\longrightarrow }%
G_{1}\longrightarrow ...G_{h}\overset{c_{h}\xi _{h},\theta _{h}}{%
\longrightarrow }G,$\smallskip

\noindent the sequence $M=c_{1},c_{2},...c_{h}$ of applied clauses will be
called \emph{template}. The whole derivation will be denoted by $G_{o}%
\overset{M,\theta }{\longrightarrow }G$, where $\theta =\theta _{1}\theta
_{2}...\theta _{h}$, or simply $G_{o}\overset{M}{\longrightarrow }G$, if the
substitution $\theta $ does not need to be pointed out. We use the notation $%
G_{o}\overset{M}{\longrightarrow }\bullet $, when there is not interest in
specifying the final resolvent. Given a template $M$, the notation $\#M$
will indicate the number of clauses in $M$. In many cases, we will consider
concatenation of templates, which is denoted by a vertical 
bar ``\TEXTsymbol{\vert}''.

It is intuitive that, given a derivation, any subset of atoms in the current
resolvent \emph{derives from} other specific atoms in preceding resolvents.
As it will be clear in the sequel, this idea plays an important role in the
development of this paper. Thus, it is convenient to give some formal
definitions. Precisely, let us consider a p-SLD derivation of the form 
$Dr=(F+G\overset{H}{\longrightarrow }Q)$. The following two intuitive
concepts will be characterised:

\begin{enumerate}
\item[a)]  the \emph{sub-resolvent of}$\;F\emph{\;}$\emph{in} $Dr$, i.e. the
subset of p-atoms in $Q$ which derive from the 
subgoal$\;F\;$(denoted by $Q/F $),\medskip

\item[b)]  the \emph{sub-template of}$\;F\emph{\;}$\emph{in} $Dr$, i.e. the
sequence of clauses which are applied to p-atoms of$\;F\;$and p-atoms
derived from $F$, extracted in the order from the template $H$ (denoted by 
$H/F$).
\end{enumerate}

\begin{definition}[sub-resolvents and sub-templates]
\label{d2.3.4}

\begin{enumerate}
\item[i)]  Given a derivation step of the following form, where 
$c=(ht\longleftarrow B)$:

\qquad $a|(F+G)\overset{c}{\longrightarrow }(Q=((F+G)+B\xi 
\underline{\pi })\alpha )$,\hfill (1)\qquad 
\smallskip

let us define \emph{sub-resolvents} and \emph{sub-templates} in (1)
as follows:

\qquad $Q/a=B\xi \underline{\pi }\alpha ,\;\;Q/F=F\alpha
,\;\;Q/(a|F)=Q/a+Q/F $

\qquad $c/a=c,\;\;c/F=\varnothing ,\;\;c/(a|F)=c.$\medskip

\item[ii)]  Given a derivation of the form:

\qquad $F+G\overset{c}{\longrightarrow }Q\overset{K}{\longrightarrow }R,$
\hfill (2)\qquad
\smallskip
 
let us recursively define \emph{sub-resolvents} and \emph{sub-templates} in 
(2) as follows:

\qquad $R/F=R/(Q/F)$,

\qquad $(c|K)/F=(c/F)|(K/(Q/F))$.
\end{enumerate}
\end{definition}

It is worth noting that the notation relative to sub-templates and
sub-resolvents can be ambiguous. Indeed consider:\smallskip

$G+F\overset{D}{\longrightarrow }Q\hfill $(3)\qquad \smallskip

$G+F^{\prime }\overset{D}{\longrightarrow }Q^{\prime }$.\hfill (4)\qquad
\smallskip

\noindent It is possible that $D/G$ with respect to (3) is different from $%
D/G$ with respect to (4). In the following of the paper, when such a kind of
ambiguity will possibly arise, we exploit a refined notation of evident
meaning, like $D/^{3}/G$ and $D/^{4}/G$. As an example, let us consider $G=a$%
, $F=b$, $F^{\prime }=d$ and $D=c$ such that

$G+F=a|b\overset{c}{\longrightarrow }Q\hfill $(3b)\qquad \smallskip

$G+F^{\prime }=d|a\overset{c}{\longrightarrow }Q^{\prime }.$\hfill
(4b)\qquad \smallskip

\noindent Then, $D/^{3b}/G=c$ and $D/^{4b}/G=empty$.

\subsection{\label{subs2.4}Congruent lowering of derivation steps}

This section introduces some important ideas. Precisely, the concepts of 
\emph{specialisation}, \emph{lowering}, and finally \emph{congruent lowering}
are defined and analysed. Congruent lowering is basic for the
characterisation of the general concept of scheduling rule, as well as of
the class of specialisation independent scheduling rules (see Section \ref
{s3.specFree}) to which the results about redundancy elimination tolerance
of Section \ref{s5.reduElim} refer.

Substitutions and renamings are basic concepts in Logic Programming. In
agreement with usual terminology, if a substitution is applied to a goal, an 
\emph{instance} is obtained, while, if a renaming is used, a \emph{variant}
of the original goal is produced. Goals which are equal up to renamings are
in essence equivalent goals. Practically all the results of Logic
Programming are insensible to renamings. An instance may be considered as a
specialised version of the original goal, while any goal is more general
with respect to its instances. The above concepts are easily adjusted in the
frame of priority goals. Intuitively, the application of a
renaming/substitution corresponds to the application of a
renaming/substitution together with a shifting. Actually, as it will be
clear in the following, we are interested in an idea of \emph{specialisation}
of a given p-goal which extends the traditional concept of instantiation. In
essence, we will consider couples of p-goals such that the second one is
obtained from the first one by performing in the order:\smallskip

- the application of a generic substitution $\lambda $ and a shifting 
\underline{$\sigma $},\smallskip

- the embedding in a generic context $X$ of other p-atoms.\smallskip

\begin{definition}[specialisation]
\label{d2.4.1}
A p-goal$\;F\;$is a \emph{%
specialisation} of a p-goal $a|K$ by $X$, if a shifting \underline{$\sigma $}
and a substitution $\lambda $ exist such that\smallskip

$F=a\lambda \underline{\sigma }|(K\lambda \underline{\sigma }+X)$.\medskip
\end{definition}

It is worth noting that our idea of specialisation is essentially symmetric
to the concept of subsumption by an instance (see \cite{4.BAK91}). A goal$%
\;G\;$\emph{subsumes} (as list) a goal$\;F\;$\emph{by an instance}, if a
substitution $\lambda $ exists such that $G\lambda \subseteq _{L}F$. Indeed,
considering that any shifting preserves the order of the atoms, it is
evident that, if$\;F\;$is a specialisation of $a|K$ by $X$, i.e.$%
\;F\;=a\lambda \underline{\sigma }|(K\lambda \underline{\sigma }+X)$, then $%
a|K$ subsumes (as list)$\;F\;$by the instance $(a|K)\lambda $.

The term ``lifting'' is used in Logic Programming to express that a
derivation step (or a whole derivation) which is possible from a goal $%
A\lambda $ is repeated starting from the more general goal $A$. Analogously,
we use the term lifting to mean that a derivation step (or a whole
derivation) which is possible from a specialisation of $a|K$, i.e. from a
p-goal $a\lambda \underline{\sigma }|(K\lambda \underline{\sigma }+X)$, is
repeated starting from $a|K$.
In the sequel of the paper, we will use the
dual concept of ``lowering''. In other words, the term lowering will mean
that a derivation step (or a whole derivation) from a p-goal $a|K$ is
repeated, when possible, starting from a specialisation $a\lambda \underline{%
\sigma }|(K\lambda \underline{\sigma }+X)$ of $a|K$. Then, let us give the
following definition which refers to single derivation steps.\smallskip

\begin{definition}[lowering of derivation steps]
\label{d2.4.2}
Let us consider two
priority derivation steps of the type$\;G\overset{c}{\longrightarrow }%
\bullet $ and $F\overset{c}{\longrightarrow }\bullet $. We will say that the
second step is a \emph{lowering of} the first one \emph{by} $X$, if the
p-goal$\;F\;$is a specialisation of$\;G\;$by $X$.\medskip
\end{definition}

Let us consider two derivation steps (Ds1) and (Ds2), such that (Ds2) is a
lowering of (Ds1) by $X$, and let $c=(ht\longleftarrow B)$. By definition of
derivation step, they have the following form:\medskip

$a|K\overset{c}{\longrightarrow }(K+B\xi ^{\prime }\underline{\theta }
^{\prime })\alpha ^{\prime }\hfill $(Ds1)\qquad \medskip

$a\lambda \underline{\sigma }|(K\lambda \underline{\sigma }+X)\overset{c}{
\longrightarrow }(X+K\lambda \underline{\sigma }+B\xi ^{\prime \prime }
\underline{\theta }^{\prime \prime })\alpha ^{\prime \prime }.\hfill $
(Ds2)\qquad \medskip

The definition of lowering of derivation steps does not impose any
similarity in the way priority values are handled in couples of derivation
steps like (Ds1) and (Ds2). In particular, no analogy is required about the
positions new atoms go to occupy with respect to old ones in the resolvents
produced by (Ds1) and (Ds2). Indeed the shifting $\underline{\theta }
^{\prime }$ and $\underline{\theta }^{\prime \prime }$ are completely
independent, so that the positions of atoms of $B\xi ^{\prime \prime }
\underline{\theta }^{\prime \prime }$, with respect to atoms of $K\lambda 
\underline{\sigma }$, will be in general different from the positions
occupied by atoms of $B\xi ^{\prime }\underline{\theta }^{\prime }$ with
respect to atoms of $K$. Nevertheless, in the rest of the paper special
importance will be given to derivation step lowering such that the
positioning of new atoms, with respect to the old ones in $K$ and $K\lambda 
\underline{\sigma }$, is maintained passing from (Ds1) to (Ds2). In such
hypothesis, we will say that the lowering is a \emph{congruent lowering}.
\medskip

As an elementary example, let us consider a clause like $c=a\longleftarrow
b_{1}|b_{2}$ and the following derivation steps, such that (2) is a lowering
of (1) by $x_{1}|x_{2}$:\smallskip

$a|k_{1}|k_{2}\overset{c}{\longrightarrow }b_{1}^{\prime }\underline{\theta }%
^{\prime }|k_{1}|b_{2}\underline{\theta }^{\prime }|k_{2}\hfill $(1)\qquad
\medskip

$a|x_{1}|k_{1}|x_{2}|k_{2}\overset{c}{\longrightarrow }x_{1}|b_{1}\underline{%
\theta }^{\prime \prime }|k_{1}|x_{2}|b_{2}\underline{\theta }^{\prime
\prime }|k_{2}\hfill $(2)\qquad \medskip

\noindent In (1) and (2) the relative positions of atoms $b_{1}$ and $b_{2}$ with
respect to $k_{1}$ and $k_{2}$ are the same, then (2) is a congruent
lowering of (1). Now, let us consider the following other derivation step
(3):\medskip

$a|x_{1}|k_{1}|x_{2}|k_{2}\overset{c}{\longrightarrow }x_{1}|k_{1}|b_{1}%
\underline{\tau }|x_{2}|b_{2}\underline{\tau }|k_{2}\hfill $(3)\qquad
\smallskip

\noindent Also (3) is a lowering of (1) by $x_{1}|x_{2}$. However, in this case the
positioning of atoms $b_{1}$ and $b_{2}$ with respect to $k_{1}$ and $k_{2}$
is not maintained passing from (1) to (3), so that (3) is not a congruent
lowering of (1). Variable substitutions are not considered in the above
examples. Indeed, in agreement with the following formal Definition \ref
{d2.4.3}, they are not really influent for a lowering to be congruent or not.
\smallskip

\begin{definition}[congruent lowering]
\label{d2.4.3}
Let us consider two derivation
steps of the form (Ds1) and (Ds2) above, i.e. two derivation
steps such that the second one is a lowering of the first one by $X$. We
will say that step (Ds2) is a \emph{congruent lowering} of step 
(Ds1) \emph{by} $X$ if a shifting \underline{$\rho $} exists with:\smallskip

$K\underline{\rho }=K\underline{\sigma }$ and $B\underline{\theta }^{\prime }%
\underline{\rho }=B\underline{\theta }^{\prime \prime }$.\hfill 
(c1)\qquad \medskip
\end{definition}

It is apparent that the desired analogy, in positioning new atoms in the two
derivation steps (Ds1) and (Ds2), is imposed by means of condition (c1)
above in Definition \ref{d2.4.3}. Indeed, condition (c1) says that the
shifting \underline{$\rho $} creates a correspondence between atoms of $K+B%
\underline{\theta }^{\prime }$ and atoms of $K\underline{\sigma }+B%
\underline{\theta }^{\prime \prime }$, such that old atoms are mapped in old
atoms (see $K\underline{\rho }=K\underline{\sigma }$) and new atoms in new
ones (see $B\underline{\theta }^{\prime }\underline{\rho }=B\underline{%
\theta }^{\prime \prime }$). Since any shifting maintains atom precedence,
it is intuitive that congruent allocation of new atoms is imposed. More
specifically, let us consider the generic atom $b$ of$\;B\;$and
assume:\smallskip

$K+B\underline{\theta }^{\prime }=M^{\prime }|b\underline{\theta }^{\prime
}|N^{\prime }$,\smallskip

$K\underline{\sigma }+B\underline{\theta }^{\prime \prime }=M^{\prime \prime
}|b\underline{\theta }^{\prime \prime }|N^{\prime \prime }$.\smallskip

\noindent It is immediate to verify that$^{[}$\footnote{%
The notation $"K\underline{\sigma }+B\underline{\theta }"=^{(c1)}K\underline{%
\rho }+B\underline{\theta }^{\prime }\underline{\rho }"$ expresses that the
formula (c1) must be used to establish the equality. Similar advising will
be used frequently in the sequel.}$^{]}$:\smallskip

$M^{\prime \prime }|b\underline{\theta }^{\prime \prime }|N^{\prime \prime
}=K\underline{\sigma }+B\underline{\theta }^{\prime \prime }=^{(c1)}K%
\underline{\rho }+B\underline{\theta }^{\prime }\underline{\rho }=(K+B%
\underline{\theta }^{\prime })\underline{\rho }=\smallskip $

$\hfill =(M^{\prime }|b\underline{\theta }^{\prime }|N^{\prime })\underline{%
\rho }=M^{\prime }\underline{\rho }|b\underline{\theta }^{\prime }\underline{%
\rho }|N^{\prime }\underline{\rho }.\qquad $\smallskip

Now, by $B\underline{\theta }^{\prime }\underline{\rho }=B\underline{\theta }%
^{\prime \prime }$ in (c1) and Ax-iv in Property \ref{py2.2.1}, we have that 
$b\underline{\theta }^{\prime }\underline{\rho }=b\underline{\theta }%
^{\prime \prime }$. Then, by Property \ref{py2.1.1}-ii) it is $M^{\prime }%
\underline{\rho }=M^{\prime \prime }$, and then also $\#M^{\prime
}=\#M^{\prime }\underline{\rho }=\#M^{\prime \prime }=n$, for $n$ positive
integer. In essence, considered the generic atom $b$ of $B$, it is found in
the $(n+1)^{th}$ position in $K+B\underline{\theta }^{\prime }$ as well as
in $K\underline{\sigma }+B\underline{\theta }^{\prime \prime }$. In other
words, new atoms from$\;B\;$are positioned in (Ds1) with respect to old ones
(i.e. atoms of $K$) exactly as it happens in (Ds2) with respect to $K%
\underline{\sigma }$. It is evident that the presence of various
substitutions in (Ds1) and (Ds2) does not interfere with the above
positional considerations.

\begin{example}[lowering and congruent lowering]
\label{e2.4.1} 

\noindent Let us consider a clause of the form$\;c\;=(a\longleftarrow q[1])$
and the two following derivation steps:\smallskip

$a[2]|\{b[\mathbf{3}]\}\overset{c}{\longrightarrow }\{b[\mathbf{3}],q[%
\underline{10}]\}$,\hfill (1)\qquad \smallskip

$a[9]|\{b[12],b[13],d[15]\}\overset{c}{\longrightarrow }%
\{b[12],q[12.5],b[13],d[15]\}$.\hfill (2)\qquad \smallskip

\noindent In step (1), old atoms are pointed out in bold and new ones are
underlined.

\begin{enumerate}
\item[a)]  In agreement with Definition \ref{d2.4.3}, step (2) is a lowering
of (1) by $X=\{b[13],d[15]\}$, with $K\underline{\sigma }=\{b[12]\}$.
Pointing out old and new atoms, derivation step (2) can be written as
follows:\smallskip

$a[9]|\{b[\mathbf{12}],b[13],d[15]\}\overset{c}{\longrightarrow }\{b[\mathbf{%
12}],q[\underline{12.5}],b[13],d[15]\}$.\smallskip 

It is evident that (2) is a congruent lowering of (1) by $X$, with any
shifting \underline{$\rho $} such that $\underline{\rho }\supseteq
\{3/12,10/12.5\}$.\smallskip 

\item[b)]  Step (2) is a lowering of (1) also by $X^{\prime
}=\{b[12],d[15]\} $, with $K\underline{\sigma }^{\prime }=\{b[13]\}$.
However (2) is not a congruent lowering of (1) by $X^{\prime }$. In fact, in
agreement with this second viewpoint, derivation step (2) can be written as
follows:\smallskip

$a[9]|\{b[12],b[\mathbf{13}],d[15]\}\overset{c}{\longrightarrow }\{b[12],q[%
\underline{12.5}],b[\mathbf{13}],d[15]\}$.\smallskip

As a consequence, for step (2) being a congruent lowering of step (1) by $%
X^{\prime }$, a shifting $\underline{\rho }^{\prime }$ might exist such that 
$\underline{\rho }^{\prime }\supseteq \{3/13,10/12.5\}$, which is not an
increasing function.~$\square \qquad $\smallskip 
\end{enumerate}
\end{example}

We close this section considering a couple of p-goals$\;F\;$and$\;G\;$such
that they are specialisations of each other, i.e.$\;F\;$is a specialisation
of$\;G\;$by a subgoal $X$ and$\;G\;$is a specialisation of$\;F\;$by $Y$. In
this case it must be$\;F\;=G\lambda \underline{\sigma }+X$ and$\;G\;=F\tau 
\underline{\rho }+Y$, which yields:\smallskip

$G\;=F\tau \underline{\rho }+Y=(G\lambda \underline{\sigma }+X)\tau 
\underline{\rho }+Y=G\lambda \tau \underline{\sigma }\underline{\rho }+X\tau 
\underline{\rho }+Y$.\smallskip

As a consequence $\lambda $ must be a renaming for$\;G\;$and $%
X=Y=\varnothing $ must hold, which means that$\;F\;=G\lambda \underline{%
\sigma }$ where $\lambda $ is a renaming. It is evident that the relation ``$%
F=G\lambda \underline{\sigma }$, for a renaming $\lambda $ and a shifting $%
\underline{\sigma }$'' can be seen as the translation of the usual notion of
``$F$ being variant of $G$'' in the frame of p-SLD resolution. In this
sense, we will usually say that$\;F\;$is a \emph{p-variant} of a $G$, to
mean that$\;F\;$and$\;G\;$are specialisations of each other.

Analogously, two derivation steps may be \emph{lowerings of each other}, as
well as \emph{congruent lowerings of each other}. Two derivation steps $%
Ds_{1}$ and $Ds_{2}$ are lowerings of each other if the initial goals are
p-variants and the same clause is applied, i.e. it is $Ds_{1}=(A\overset{c}{%
\longrightarrow }\bullet )$ and $Ds_{2}=(A\lambda \underline{\sigma }%
\overset{c}{\longrightarrow }\bullet )$, where $\lambda $ is a renaming. Two
derivation steps are congruent lowerings of each other if they have the
form:\smallskip

$Ds_{1}=a|K\overset{c}{\longrightarrow }(K+B\xi ^{\prime }\underline{\theta }%
^{\prime })\alpha ^{\prime }$ and $Ds_{2}=(a|K)\lambda \underline{\sigma }%
\overset{c}{\longrightarrow }(K\lambda \underline{\sigma }+B\xi ^{\prime
\prime }\underline{\theta }^{\prime \prime })\alpha ^{\prime \prime }$%
,\smallskip

\noindent where $c=(ht\longleftarrow B)$, $\lambda $ is a renaming, and the
equalities $K\underline{\rho }=K\underline{\sigma }$ and $B\underline{\theta 
}^{\prime }\underline{\rho }=B\underline{\theta }^{\prime \prime }$ hold for
a shifting $\underline{\rho }$.

It is worth noting that by the preceding argument if two derivation steps
are lowerings of each other the contexts must be empty.

\subsection{\label{subs2.5}State priority scheduling rules}

Now, we use the notion of being congruent lowerings of each other to define
the ideas of \emph{determinism} and \emph{completeness} of a set of
derivation steps. Both concepts are basic for the definition of state
priority scheduling rules.

\begin{definition}[determinism]
\label{d2.5.1}
A set$\;S\;$of priority derivation steps
is \emph{deterministic} if, for each couple of derivation steps $Ds_{1}$ and 
$Ds_{2}$ in$\;S,$ the following implication holds:\smallskip

$Ds_{1}$ and $Ds_{2}$ are lowerings of each other\smallskip

$\qquad \Longrightarrow $ $Ds_{1}$ and $Ds_{2}$ are congruent lowerings of
each other.\medskip
\end{definition}

In other words, the definition of determinism imposes that two derivation
steps, which apply the same clause to p-variant initial goals, give place to
congruent allocations of new atoms. Now let us give the definition of
completeness of a set of derivation steps.

\begin{definition}[completeness]
\label{d2.5.2}
A set S of priority derivation steps is 
\emph{complete}, if the following assertions hold:\smallskip

i) $\exists $ $Ds$ derivation step of the type$\;G\overset{c}{%
\longrightarrow }\bullet ,$\smallskip

$\qquad \Longrightarrow $ $\exists $ $Ds^{\prime }$ of the type$\;G\overset{c%
}{\longrightarrow }\bullet $ with $Ds^{\prime }\in S$,\smallskip

ii) $\forall Ds^{\prime },Ds$ derivation steps with $Ds\in S$,\smallskip

\qquad $Ds^{\prime }$ and $Ds$ are congruent lowerings of each other $%
\Longrightarrow $ $Ds^{\prime }\in S$.\smallskip
\end{definition}

Assertion i) of the above definition states that, if a clause$\;c\;$is
applicable to a p-goal $G$, i.e. a derivation step exists of the type$\;G%
\overset{c}{\longrightarrow }\bullet $, the application of the clause$\;c\;$%
to$\;G\;$is indeed possible in any complete set of derivation steps.
Assertion ii) assures that$\;S\;$is closed with respect to being congruent
lowerings of each other. In other words, let $Ds^{\prime }=(G\overset{c}{%
\longrightarrow }Q)\in S$ be a derivation step, then every other $Ds^{\prime
\prime }=(F\overset{c}{\longrightarrow }R)$ must belong to$\;S,$ if$\;F\;$is
a p-variant of$\;G\;$and new atoms are allocated in$\;R\;$as it is done in $%
Q $. Now, the formal definition of state priority scheduling rules can be
easily given, by combining the properties of determinism and completeness.

\begin{definition}[state priority scheduling rules]
\label{d2.5.3} 
A \emph{state
priority scheduling rule} is a complete and deterministic set of priority
derivation steps.\medskip
\end{definition}

It can be easily verified that the leftmost selection rule, adopted by the
Prolog execution mechanism, is a state priority scheduling rule. The very
nature of a state scheduling rule is characterised by the following
Definition \ref{d2.5.4}. Indeed, the definition simply says that a p-SLD
derivation \emph{is via} a state scheduling rule$\;S\;$if all derivation
steps are admitted in the rule$\;S,$ i.e. they all belong to the set of
derivation steps which$\;S\;$is constituted by.

\begin{definition}[derivations via S]
\label{d2.5.4}

\begin{enumerate}
\item[i)]  Given a set$\;S\;$of derivation steps, the notation $\Delta (S)$
represents the whole of p-SLD derivations which are composed of derivation
steps in$\;S.$\smallskip

\item[ii)]  Given a state scheduling rule$\;S,$ the set $\Delta (S)$ is the
set of \emph{p-SLD derivations via}$\;S.$
\end{enumerate}
\end{definition}

In the sequel of the paper we only consider state priority scheduling rules,
which therefore will be called just \emph{scheduling rules}. The following
notations will be used frequently. Given a set$\;S\;$of derivation steps, a
clause$\;c\;$and a template $M$, we will denote by\smallskip

$G\overset{S,c}{\longrightarrow }R\;\;\;$and$\;\;\;G\overset{S,M}{%
\longrightarrow }R$\smallskip

\noindent the fact that the derivation step $(G\overset{c}{\longrightarrow }%
R)\in S\;$and the p-SLD derivation $(G\overset{M}{\longrightarrow }R)\in
\Delta (S)$, respectively. In the case that the exploited logic program must
be pointed out, a notation like\smallskip

$(G\overset{S,M.P}{\longrightarrow }R)$\smallskip

\noindent will be used to specify that the derivation is via$\;S\;$in the
program P, i.e. every clause of the template $M$ belongs to P. The notion of
p-SLD tree via$\;S\;$could be characterised in complete analogy with the
usual one of SLD tree.

Let us close this section with a property, which can be easily shown on the
basis of completeness and will be used several times in the sequel. Property 
\ref{py2.5.1} asserts that if a clause$\;c\;$can be applied to a p-goal $%
a\gamma |G$, every complete set of derivation steps allows$\;c\;$to be
applied to any p-goal of the form $a|F$. Since the atom $a$ is more general
than $a\gamma $, the property may also be interpreted as a sort of lifting
of derivation steps. However, the subgoals$\;G\;$and$\;F\;$are left
unrelated at all. The evident explication is that they have no active role
in rewriting operations. Moreover, the property recalls that new variables
can be always chosen so that conflicts are avoided with arbitrary
pre-established sets of variables. The formal proof of this rather intuitive
property can be found in Appendix A.

In the statement of Property \ref{py2.5.1} and in the sequel of the paper,
given a p-SLD derivation $Dr$, the notation $nvar(Dr)$ will represent the
set of standardisation apart variables which are introduced during the
derivation $Dr$. In the case of a single derivation step $Ds=(A\overset{c\xi 
}{\longrightarrow }\bullet )$, it is $nvar(Ds)=var(c\xi )$.
\medskip

\begin{property}
\label{py2.5.1}Let$\;S\;$be a complete set of derivation steps. Given two
p-goals $a\gamma \underline{\tau }|G$ and $a|F$, let us fix arbitrarily a
finite set $V$ of variables. The following implication holds:\smallskip

$\exists Ds$ derivation step of the type $a\gamma \underline{\tau }|G%
\overset{c}{\longrightarrow }\bullet $\medskip

$\qquad \Longrightarrow \;\;$\ $\exists Ds^{\prime }$ of the type $a|F%
\overset{c}{\longrightarrow }\bullet ,$ with $Ds^{\prime }\in S\;$and $%
nvar(Ds^{\prime })\cap V=\varnothing $.
\end{property}
\medskip

\section{\label{s3.specFree}Specialization independent scheduling rules}

Now, we will exploit the notion of congruent lowering in order to introduce
the concept of \emph{specialisation independence}. This concept will be used
to characterise the class of scheduling rules that are the main object of
the paper (\emph{specialisation independent scheduling rules}). In fact, all
our results for termination and loop check completeness preserving will
refer to such a class of scheduling rules. In Section \ref{s4.stacQueu}, a
second characterisation of the same class is given which has an operational
nature and is surprisingly different in appearance.

The definition of \emph{specialisation independence} enforces the idea of
determinism. Indeed, in agreement with the Definition \ref{d3.1} below,
every lowering is required to be a congruent lowering. In other words, the
congruence in the allocation of new atoms must hold any time the initial
goals of two derivation steps are related by specialisation and the same
clause is used. This can be interpreted saying that the positioning of new
atoms with respect to old ones is \emph{independent of goal specialisation},
which means independent of goal instantiation as well as of the addition of
a group $X$ of other atoms.
\bigskip

\begin{definition}[specialisation independence]
\label{d3.1}
A set$\;S\;$of priority
derivation steps is \emph{specialisation independent} if, for every couple
of steps $Ds_{1}$ and $Ds_{2}$ in$\;S,$ the following implication
holds:\smallskip

$Ds_{2}$ is a lowering of $Ds_{1}$ by $X$\smallskip

$\qquad \Longrightarrow \;\;\;Ds_{2}$ is a congruent lowering of $Ds_{1}$ by 
$X$.
\end{definition}
\bigskip

\begin{definition}[specialisation independent scheduling rules]
\label{d3.2}
A \emph{%
specialisation independent scheduling rule} is a complete and specialisation
independent set of priority derivation steps.
\end{definition}
\bigskip

In the next two sections, we provide some results about p-SLD derivations
via specialisation independent scheduling rules. The results will be
frequently exploited in the sequel.

\subsection{\label{subs3.1}Derivation lowering}

In this section we give results which relate resolvents coming from a couple
of derivation steps in the congruent lowering relationship. Then, by Lemma 
\ref{L3.1.1}, the analysis is extended to couples of whole derivations,
developed via specialisation independent scheduling rules. We start by
presenting a preliminary statement (Property \ref{py3.1.1}) which holds for
every couple of derivation steps that are in the lowering relationship. In
reference to derivation steps (1) and (2) below, the preliminary property
says that, if we abstract from atom positioning and ignore the additional
subgoal $X$, the resolvent of (2) is an instance of the resolvent of (1).
Property \ref{py3.1.1} can be shown following the line exploited for proving
the Variant Lemma (see \cite{1.A90}), which is done in Appendix A for the
sake of completeness of the paper.
\medskip

\begin{property}
\label{py3.1.1}Let $c=(ht\longleftarrow B)$ be a clause. Let us consider two
derivation steps like (1) and (2), where (2) is a
lowering of (1) by $X$. The following implication holds:\smallskip

$a|K\overset{c\xi ^{\prime }}{\longrightarrow }(K+B\xi ^{\prime }\underline{%
\theta }^{\prime })\mu ^{\prime }$,\hfill (1)\qquad \smallskip

$a\tau \underline{\sigma }|(K\tau \underline{\sigma }+X)\overset{c\xi
^{\prime \prime }}{\longrightarrow }(K\tau \underline{\sigma }+B\xi ^{\prime
\prime }\underline{\theta }^{\prime \prime }+X)\mu ^{\prime \prime }$\hfill 
(2)\qquad \smallskip

$\qquad \Longrightarrow \;\exists \delta $ such that $K\tau \mu ^{\prime
\prime }=K\mu ^{\prime }\delta $ and $B\xi ^{\prime \prime }\mu ^{\prime
\prime }=B\xi ^{\prime }\mu ^{\prime }\delta $,\smallskip

\qquad \qquad\ where $\delta $ is a renaming, if $\tau $ is a renaming.
\end{property}
\bigskip

Property \ref{py3.1.2} completes Property \ref{py3.1.1}, taking into account
the preservation of atom scheduling in the case of congruent lowering. It
states that, if we ignore the additional subgoal $X$, resolvents are
preserved up to a substitution and a shifting. In reference to derivation
steps (1) and (2) below, this means that, apart from $R/X$, the resolvent$%
\;R\;$in (2) is an instance of $Q$ such that also atom scheduling is
maintained.

\begin{property}
\label{py3.1.2}Let $c=(ht\longleftarrow B)$ be a clause. Let us consider two
derivation steps of the type (1) and (2), such that the second
one is a congruent lowering of the first one by $X$:\smallskip

$a|K\overset{c}{\longrightarrow }Q$,\hfill (1)\qquad \smallskip

$a\tau \underline{\pi }|(K\tau \underline{\pi }+X)\overset{c}{%
\longrightarrow }R$.\hfill (2)\qquad \smallskip

\noindent The following assertion holds:\smallskip

$\exists \delta ,\underline{\rho }$ such that $R/((a|K)\tau \underline{\pi }%
)=Q\delta \underline{\rho }$,\smallskip

where $\delta $ is a renaming if $\tau $ is a renaming.
\end{property}

\begin{proof}
Let $c=(ht\longleftarrow B)$, so that $Q$ and$\;R\;$%
may be written as follows:\smallskip

$Q=(K+B\xi ^{\prime }\underline{\theta }^{\prime })\mu ^{\prime }$,\smallskip

$R=(K\tau \underline{\pi }+X+B\xi ^{\prime \prime }\underline{\theta }%
^{\prime \prime })\mu ^{\prime \prime }$.\smallskip

\noindent Since step (2) is a congruent lowering of (1) by $X$, a shifting 
\underline{$\rho $} exists such that:\smallskip

$K\underline{\rho }=K\underline{\pi },\;\;\;B\underline{\theta }^{\prime }%
\underline{\rho }=B\underline{\theta }^{\prime \prime }$.\hfill (3)\qquad
\smallskip

\noindent By definition of sub-resolvent and (3), we have:\smallskip

$R/((a|K)\tau \underline{\pi })=B\xi ^{\prime \prime }\underline{\theta }%
^{\prime \prime }\mu ^{\prime \prime }+K\tau \underline{\pi }\mu ^{\prime
\prime }=^{(3)}B\xi ^{\prime \prime }\mu ^{\prime \prime }\underline{\theta }%
^{\prime }\underline{\rho }+K\tau \mu ^{\prime \prime }\underline{\rho }.$%
\hfill (4)\qquad \smallskip

\noindent Now, we apply Property \ref{py3.1.1} to (1) and (2), deriving that
a substitution $\delta $ exists such that:\smallskip

$K\tau \mu ^{\prime \prime }=K\mu ^{\prime }\delta $ \ \ and \ $B\xi
^{\prime \prime }\mu ^{\prime \prime }=B\xi ^{\prime }\mu ^{\prime }\delta ,$%
\hfill (5)\qquad \smallskip

where $\delta $ is a renaming if $\tau $ is a renaming.

\noindent As a consequence, we have that:\smallskip

$R/((a|K)\tau \underline{\pi })=^{(4)}B\xi ^{\prime \prime }\mu ^{\prime
\prime }\underline{\theta }^{\prime }\underline{\rho }+K\tau \underline{\rho 
}\mu ^{\prime \prime }=^{(5)}B\xi ^{\prime }\mu ^{\prime }\delta \underline{%
\theta }^{\prime }\underline{\rho }+K\mu ^{\prime }\delta \underline{\rho }%
=Q\delta \underline{\rho },$\smallskip

where $\delta $ is a renaming if $\tau $ is a renaming.
\end{proof}
\bigskip 

The following Lemma \ref{L3.1.1} may be seen as the extension of Property 
\ref{py3.1.2} to whole derivations, provided that the used scheduling rule
is specialisation independent. Note that, given a derivation like (1) in the
statement below, if a derivation like (2) exists, it can be considered as a
lowering of (1). Indeed, the initial p-goal $X+G\gamma \underline{\tau }$ is
a specialisation of$\;G\;$by $X$, and the sequence $E$ of clauses is applied
in the same order to atoms deriving from $G\gamma \underline{\tau }$ in
derivation (2). In this sense we will regard Lemma \ref{L3.1.1} as a
``specialisation independent lowering lemma''.

\begin{lemma}[specialisation independent lowering lemma]
\label{L3.1.1}
Let$\;S\;$%
be a specialisation independent scheduling rule and consider two p-SLD
derivations like (1) and (2). The following implication
holds:\smallskip

$G\overset{S,E}{\longrightarrow }Q$,\hfill (1)\qquad \smallskip

$G\gamma \underline{\tau }+X\overset{S,D}{\longrightarrow }R$, with $%
D/(G\gamma \underline{\tau })=E$\hfill (2)\qquad \smallskip

$\;\;\Longrightarrow \;\exists \sigma ,\underline{\rho }$ such that $%
R/(G\gamma \underline{\tau })=Q\sigma \underline{\rho }$,

\qquad\ \ where $\sigma $ is a renaming if $\gamma $ is a renaming and $%
D/X=\varnothing $.\hfill (p1)\qquad \smallskip
\end{lemma}

\begin{proof}
Let us first prove the thesis, apart from the fact
(p1). The proof is by induction on the length of $D$. If $\#D$ is equal to
zero, the thesis is trivially true. Let us suppose that $\#D$ is greater
than zero. Two different cases must be considered, i.e. the first clause of $%
D$ (say $c$) is applied either to an atom of $X$ or to an atom of $G\gamma 
\underline{\tau }$.\medskip

\noindent \underline{First case} (\emph{The clause}$\;c\;$\emph{is applied
to an atom of }$X$).

\noindent In this case derivation (2) may be rewritten as:\smallskip

$X+G\gamma \underline{\tau }\overset{S,c\eta ,\alpha }{\longrightarrow }%
G\gamma \underline{\tau }\alpha +X^{\prime }\overset{S,D^{\prime }}{%
\longrightarrow }R$,\hfill (3)\qquad \smallskip

with $D^{\prime }/(G\gamma \underline{\tau }\alpha )=D/(G\gamma \underline{%
\tau })=E$.\smallskip

\noindent By inductive hypothesis, applied to the tail of derivation (3) and
derivation (1), we have:\smallskip

$\exists \sigma ,\underline{\rho }$ such that $R/(G\gamma \underline{\tau }%
)=R/(G\gamma \underline{\tau }\alpha )=^{(ind.hyp.)}Q\sigma \underline{\rho }%
.\medskip $

\noindent \underline{Second case} (\emph{The clause}$\;c\;$\emph{is applied
to an atom of }$G\gamma \underline{\tau }$).

\noindent In in this case derivations (1) and (2) may be rewritten as (4)
and (5), respectively:\smallskip

$G\overset{S,c\xi }{\longrightarrow }Y\overset{S,E^{\prime }}{%
\longrightarrow }Q$\hfill (4)\qquad \smallskip

$X+G\gamma \underline{\tau }\overset{S,c\eta ,\alpha }{\longrightarrow }%
X\alpha +Z\overset{S,D^{\prime }}{\longrightarrow }R$\hfill (5)\qquad
\smallskip

with $D^{\prime }/Z=E^{\prime }$, $\ \ c/^{4}/G=c/^{5}/G\gamma \underline{%
\tau }=c$.\hfill (6)\qquad \smallskip

\noindent Since$\;S\;$is specialisation independent, the first step of (5)
is a congruent lowering of the first one of (4) by $X$. Then, by Property 
\ref{py3.1.2}, we have:\smallskip

$\exists \sigma ^{\prime },\underline{\rho }^{\prime }$ such that $%
Z=(X\alpha +Z)/(G\gamma \underline{\tau })=^{(Prop.\ref{py3.1.2})}Y\sigma
^{\prime }\underline{\rho }^{\prime }$.\hfill (7)\qquad \smallskip

\noindent As a consequence, recalling the first fact in (6), the inductive
hypothesis can be applied to the tails of derivations (4) and (5). Then, we
have:\smallskip

$\exists \sigma ,\underline{\rho }$ such that $R/Z=Q\sigma \underline{\rho }$.
\hfill (8)\qquad \smallskip

\noindent In conclusion, we have that:\smallskip

$R/(G\gamma \underline{\tau })=R/Z=Q\sigma \underline{\rho }$.\smallskip

\noindent In order to show the fact (p1), i.e. $\sigma $ is a renaming if $%
\gamma $ is a renaming and $D/X=\varnothing $, it is sufficient to note
that:\smallskip

- the ``first case'' does not occur at all,

- the substitutions $\sigma ^{\prime }$ and $\sigma $, mentioned in (7) and
(8), are renamings.
\end{proof}
\bigskip

The following example shows that the hypothesis of specialisation
independence is crucial for the validity of Lemma \ref{L3.1.1}.

\begin{example}
\label{e3.1.1} \ \ 
\end{example}

\noindent Let us consider a scheduling rule$\;S\;$such that new atoms are
positioned in the centre of the old resolvent. New atoms are positioned
immediately before the centre if the length of the resolvent (the rewritten
atom excluded) is odd. It is easy to recognise that Lowering Lemma \ref
{L3.1.1} does not hold for such a rule. Indeed, let P be the following
program:\smallskip

$c1=p(x)\longleftarrow q(x)[1]$\smallskip

$c2=s\longleftarrow p(b)[1]$.\smallskip

\noindent Now, in reference to the statement of Lemma \ref{L3.1.1},
let:\smallskip

$G\;=s[1],p(a)[2]$\smallskip

$G\gamma \underline{\tau }=s[1],p(a)[1.5]$ \ \ and \ $X=r[2].$\smallskip

\noindent The following are two derivations of$\;G\;$in P and $(G\gamma 
\underline{\tau }+X)$ in P, respectively:\smallskip

$\{s[1],p(a)[2]\}\overset{S,c2}{\longrightarrow }\{p(b)[1],p(a)[2]\}\overset{%
S,c1}{\longrightarrow }(\{q(b)[1],p(a)[2]\}=Q)$\smallskip

$\{s[1],p(a)[1.5],r[2]\}\overset{S,c2}{\longrightarrow }%
\{p(a)[1.5],p(b)[1.7],r[2]\}\overset{S,c1}{\longrightarrow }$\smallskip

$\hfill (\{p(b)[1.7],q(a)[1.8],r[2]\}=R).\qquad $\smallskip

\noindent Thus, no $\sigma $ and $\underline{\rho }$ can exist such
that:\smallskip

$R/(G\gamma \underline{\tau })=\{q(a)[1.8],p(b)[1.7]\}=\{q(b)[1],p(a)[2]\}%
\sigma \underline{\rho }=Q\sigma \underline{\rho }.$\smallskip

\noindent Note that $R/(G\gamma \underline{\tau })$ and $Q$ are essentially
different, even if they are considered as multisets abstracting from
priority values. It is easy to check that the used scheduling rule is not
specialisation independent, in agreement with Definition \ref{d3.2}.$\hfill
\square \qquad $

\subsection{\label{subs3.2}Derivation lifting and combining}

The following Lemma \ref{L3.2.1} is a result about p-SLD derivation lifting
which is valid for specialisation independent scheduling rules. In reference
to derivation (1) below, the lemma asserts that the sub-template of clauses,
applied to the part $G\gamma \underline{\tau }$ of the initial p-goal $%
(X+G\gamma \underline{\tau })$ in (1), can be applied again in the order
starting from the more general goal $G$, via the same scheduling rule. The
lemma also recalls that standardisation apart variables can be chosen in
order to avoid conflicts with any fixed finite set of variables. The lemma
does not relate resolvents. Indeed, Lemma \ref{L3.1.1} can be exploited to
this purpose.

\begin{lemma}[specialisation independent lifting lemma]
\label{L3.2.1}
Let$\;S\;$%
be a specialisation independent scheduling rule. Given any finite set $V$ of
variables, the following implication holds:\smallskip

$X+G\gamma \underline{\tau }\overset{S,D}{\longrightarrow }\bullet $\hfill 
(1)\qquad \smallskip

$\qquad \Longrightarrow \;\;\exists Dr=(G\overset{S,D/G\gamma \underline{%
\tau }}{\longrightarrow }\bullet )$, \ \ with \ $nvar(Dr)\cap V=\varnothing $%
.
\end{lemma}

\begin{proof}
The proof is by induction on the length of the
template $D$. If $\#D$ is zero, the assert is evident. Let us suppose that $%
\#D>0$. Two cases must be considered, i.e. either the first clause in $D$
(say $c$) is applied to an atom of $X$ or the clause$\;c\;$is applied to an
atom of $G\gamma \underline{\tau }$.\medskip

\noindent \underline{First case} (\emph{The clause}$\;c\;$\emph{is applied
to an atom of} $X$).\smallskip

\noindent Derivation (1) may be rewritten as:\smallskip

$X+G\gamma \underline{\tau }\overset{S,c\eta ,\beta }{\longrightarrow }%
X^{\prime }+G\gamma \underline{\tau }\beta \overset{S,D^{\prime }}{%
\longrightarrow }\bullet .$\hfill (2)\qquad \smallskip

\noindent By inductive hypothesis applied to the tail of (2), for any finite
set $V$ of variables, a derivation $Dr$ exists such that:\smallskip

$Dr=(G\overset{S,D^{\prime }/G\gamma \underline{\tau }\beta }{%
\longrightarrow }\bullet )$, \ \ with \ $nvar(Dr)\cap V=\varnothing $%
.\smallskip

\noindent But, by construction of (2), it is $D^{\prime }/G\gamma \underline{%
\tau }\beta =D/G\gamma \underline{\tau }$, so that the thesis is
verified.\medskip

\noindent \underline{Second case} (\emph{The clause}$\;c\;$\emph{is applied
to an atom of }$G\gamma \underline{\tau }$).\smallskip

\noindent Derivation (1) may be rewritten as follows:\smallskip

$X+G\gamma \underline{\tau }\overset{S,c\eta ,\beta }{\longrightarrow }%
X\beta +G^{\prime }\overset{S,D^{\prime }}{\longrightarrow }\bullet $,\hfill
(3)\qquad \smallskip

where $c|(D^{\prime }/G^{\prime })=D/G\gamma \underline{\tau }$.\hfill
(4)\qquad \smallskip

\noindent Let$\;G\;=a|Z$, that is $X+G\gamma \underline{\tau }=a\gamma 
\underline{\tau }|(X+Z\gamma \underline{\tau })$. By (3) and Property \ref
{py2.5.1}, we can assert that a derivation step exists like:\smallskip

$Ds^{\prime }=((G=a|Z)\overset{S,c}{\longrightarrow }R^{\prime }),$\hfill
(6a)\qquad \smallskip

with $nvar(Ds^{\prime })\cap V=\varnothing $.\hfill (6b)\qquad \smallskip

\noindent Since, by hypothesis$\;S\;$is specialisation independent, the
first step of derivation (3) is a congruent lowering of step (6a) by $X$. As
a consequence, by Property \ref{py3.1.2}, a substitution $\pi $' and a
shifting $\underline{\rho }$' exist with:\smallskip

$G^{\prime }=(X\beta +G^{\prime })/(G\gamma \underline{\tau })=R^{\prime
}\pi ^{\prime }\underline{\rho }^{\prime }.$\hfill (7)\qquad \smallskip

\noindent Then, by inductive hypothesis applied to the tail of (3), we may
assert that, a derivation $Dr^{\prime \prime }$ exists:\smallskip

$Dr^{\prime \prime }=(R^{\prime }\overset{S,D^{\prime }/G^{\prime }}{%
\longrightarrow }\bullet )$\hfill (8a)\qquad \smallskip

with $nvar(Dr^{\prime \prime })\cap (nvar(Ds^{\prime })\cup var(G)\cup
V)=\varnothing $.\hfill (8b)\qquad \smallskip

\noindent So, derivation (8a) is standardised apart with respect to (6a).
Since$\;S\;$is a state scheduling rule, (6a) and (8a) can be combined in
order to give place to an unique derivation $Dr$ such that:\smallskip

$Dr=(G\overset{c}{\longrightarrow }R^{\prime }\overset{D^{\prime }/G^{\prime
}}{\longrightarrow }\bullet )\in \Delta (S)$,\smallskip

\noindent where, by (6b) and (8b), we have also that:\smallskip

$nvar(Dr)\cap V=(nvar(Ds^{\prime })\cup nvar(Dr^{\prime \prime }))\cap
V=\varnothing .$\smallskip

\noindent By (4), the thesis is proven.
\end{proof}
\bigskip 

It is worth noting that Lowering Lemma \ref{L3.1.1} and Lifting Lemma \ref
{L3.2.1} consider couples of p-goals in a specialisation relationship, i.e.
p-goals of the form$\;G\;$and $(G\gamma \underline{\tau }+X)$. The
distinctive point is that a group $X$ of additional atoms may be present in
the second p-goal, besides the instantiation of$\;G\;$by $\gamma $. The
correspondence is obvious with the fact that Definition \ref{d3.1} requires
that positioning of new atoms is independent of goal specialisation. As it
will be clear in the following, this kind of independence is basic in order
to assure tolerance to redundancy elimination.

In \cite{7.GLM96} a class of selection rules is introduced for which
independence of atom choices from goal instantiation is assured. These rules
are named \emph{skeleton selection rules}. Indeed, they are sensible only to
a specific structural extract (the skeleton) of the applied clauses and the
initial goal in the story of a derivation. As shown in \cite{7.GLM96},
instantiation independence is sufficient to proof a Strong Lifting Lemma
which asserts that, for any skeleton rule$\;S,$ an SLD derivation of a goal $%
G\gamma $ via$\;S\;$can be lifted to a derivation of$\;G\;$via the same rule$%
\;S,$ relating in a quite strong sense the mgu's and the resolvents. On the
other hand, instantiation independence seems not sufficient to assure
redundancy elimination tolerance. For example, in agreement to the
definition in \cite{7.GLM96}, the selection rule of Example \ref{e1.1.1} is
a skeleton rule, because choices only depend on the length of the initial
goal and the ones of applied clauses. Really, choices are performed on the
unique basis of the length of the actual resolvent, so that the rule of
Example \ref{e1.1.1} can be seen as a case of $state$ skeleton selection
rule. Anyhow, the rule is not tolerant to redundancy elimination.

In order to point out the role of the hypothesis of specialisation
independence with respect to derivation lifting, let us give the following
example where Lifting Lemma \ref{L3.2.1} does not hold. Note that the used
scheduling rule is instantiation independent, but it is not specialisation
independent.

\begin{example}
\label{e3.2.1} 

\noindent Let us consider again the scheduling rule of Example \ref{e3.1.1}.
It is easy to recognise that Lifting Lemma \ref{L3.2.1} does not hold for
such a rule. Indeed, let P be the following program:\smallskip

$c1=p\longleftarrow p[1],r[2],r[3]$\smallskip

$c2=r\longleftarrow $.\smallskip

\noindent Now, in reference to the statement of Lemma \ref{L3.2.1},
let:\medskip

$G\;=\{p[1],s[2],s[3]\}$\smallskip

$G\gamma \underline{\tau }=\{p[1.1],s[2],s[2.5]\}$ \ \ and $
X=\{r[1.5],r[1.6]\}$.\medskip

\noindent In Figure \ref{f2} an infinite p-SLD derivation 
of $(G\gamma \underline{\tau }+X)$ in P is shown . 

\begin{figure}[h]
\raggedright \qquad \qquad \qquad \qquad $\{p[1.1],r[1.5],r[1.6],
s[2],s[2.5]\} \overset{S,c1}{\longrightarrow }$
\smallskip \\

\qquad \qquad \qquad \qquad $\{r[1.5],r[1.6],p[1.7],
r[1.8],r[1.9],s[2],s[2.5]\}
\overset{S,(c2,c2)}{\longrightarrow }$\medskip \\

\qquad \qquad \qquad \qquad $\{p[1.7],r[1.8],
r[1.9],s[2],s[2.5]\}$\smallskip \\ 

\qquad \qquad \qquad \qquad ....................................
\caption{} \label{f2}
\end{figure}

\noindent On the contrary, the 
only p-SLD derivation of $G$ in P is the following one\smallskip

\{$p[1],s[2],s[3]\}\overset{S,c1}{\longrightarrow }
\{s[2],p[2.5],r[2.6],r[2.7],s[3]\}.$\smallskip

\noindent which fails at the second resolvent.~$\square$
\end{example}
\bigskip

From the proofs of Lemmata \ref{L3.1.1} and \ref{L3.2.1}, the proof of two
corresponding assertions can be easily drawn. They are given in Lemma \ref
{L3.2.2} below, and are valid for all scheduling rules in the case of two
p-SLD derivations which are lowerings of each other. Part a) of the lemma
may be viewed as a form of Variant Lemma.

\begin{lemma}[determinism lemma]
\label{L3.2.2}
Let$\;S\;$be any scheduling rule
and $V$ any arbitrary finite set of variables. Then let$\;G\;$and $G^{\prime
}$ be two p-goals such that $G^{\prime }$ is a p-variant of $G$. The
following implications hold:\smallskip

a)$\;\;G\overset{S,D}{\longrightarrow }Q$ \ \ and \ $G^{\prime }\overset{S,D%
}{\longrightarrow }R$\smallskip

$\qquad \Longrightarrow \;\ \ R\;$is a p-variant of $Q$,\smallskip

b)$\;\;G^{\prime }\overset{S,D}{\longrightarrow }\bullet $\smallskip

$\qquad \Longrightarrow $ $\ \ \exists Dr=(G\overset{S,D}{\longrightarrow }%
\bullet )$, \ \ with \ $nvar(Dr)\cap V=\varnothing .\smallskip $
\end{lemma}

\begin{proof}
Let us consider part a) of the lemma. By
definition of p-variant it is $G^{\prime }=G\gamma \underline{\tau }$, for a
renaming $\gamma $ and a shifting $\underline{\tau }$. By fact (p1) in Lemma 
\ref{L3.1.1}, i.e. ``where $\sigma $ is a renaming if $\gamma $ is a
renaming and $D/X=\varnothing $'', the result appears as an immediate
consequence of the proof of Lemma \ref{L3.1.1} itself. It is sufficient to
note that, if $X$ is empty and $\gamma $ is a renaming, the fact that the
used scheduling rule is specialisation independent becomes useless. Indeed,
in reference to the proofs of Lemma \ref{L3.1.1}, though the hypothesis of
specialisation independence is dropped, the first steps of (5) and (4) are
congruent lowerings of each other, because every scheduling rule is
deterministic. Similar considerations are possible for part b) of the lemma,
in reference to the proof of Lemma \ref{L3.2.1}.
\end{proof}
\bigskip 

Now, let us give a property that is valid for all scheduling rules and
derives easily from Lemma \ref{L3.2.2}. It asserts that two p-SLD
derivations $Dr_{1}$ and $Dr_{2}$, via the same scheduling rule$\;S,$ can be
composed giving place to a longer derivation via$\;S,$ if the last resolvent
of $Dr_{1}$ coincides with the first of $Dr_{2}$.

\begin{property}[combination]
\label{py3.2.1}
Let$\;S\;$be any (state) scheduling
rule. The following implication holds:\smallskip

$\exists Dr_{1},Dr_{2}$ \ \ with \ $Dr_{1}=(G\overset{S,E}{\longrightarrow }%
F),\;\;\;Dr_{2}=(F\overset{S,H}{\longrightarrow }Q)$\smallskip

$\Longrightarrow \;\;\exists Dr=(G\overset{E}{\longrightarrow }F\overset{H}{%
\longrightarrow }R)$, \ \ with \ $Dr\in \Delta (S)$,\smallskip

\qquad\ where$\;R\;$is a p-variant of $Q$.
\end{property}

\begin{proof}
By Lemma \ref{L3.2.2}-b) applied to $Dr_{2}$, a
p-SLD derivation $Dr^{\prime }=(F\overset{S,H}{\longrightarrow }R)$ exists
with $nvar(Dr^{\prime })\cap (nvar(Dr_{1})\cup var(G))=\varnothing $. Thus, $%
Dr^{\prime }$ is standardised apart with respect to $Dr_{1}$. Since$\;S\;$is
a state scheduling rule, $Dr$ is obtained as the composition of $Dr_{1}$ and 
$Dr^{\prime }$. The fact that$\;R\;$is a p-variant of $Q$ follows from Lemma 
\ref{L3.2.2}-a), applied to $Dr_{2}$ and $Dr^{\prime }$.
\end{proof}

\section{\label{s4.stacQueu}Stack-queue selection rules}

Prolog interpreters adopt a leftmost scheduling policy such that the first
atom in the goal is always selected for rewriting and is replaced in the
resolvent by the body of the applied clause. In other words, the actual
resolvent is maintained as a \emph{stack}, the atom on the top of the stack
is always selected for rewriting, while new atoms from the applied clause
are pushed on the top of the stack. In analogy, a \emph{queue scheduling
policy} may be considered, which corresponds to a very simple case of \emph{%
fair} selection rule (see \cite{10.L87}). As for the stack scheduling policy
the first atom in the resolvent is always selected, but new atoms are
positioned at the end of the old resolvent. Thus, the resolvent is treated
as a queue of atoms and any queued atom is eventually selected in the case
of infinite derivations

In this section the class of \emph{stack-queue} scheduling rules is defined,
which is a generalisation of both stack and queue scheduling policies.
According to stack-queue rules, for any clause$\;c=(ht\longleftarrow B)$,
two p-goals $M_{s}$ and $M_{q}$ can be identified, with$\;B=M_{s}|M_{q}$,
such that the atoms in $M_{s}$ are always scheduled in \emph{stack mode}
while the atoms in $M_{q}$ are scheduled in \emph{queue mode}. More
formally, we have the following definition. As shown in the sequel of this
Section \ref{s4.stacQueu}, the stack-queue class turns out to be an
operational characterisation of the class of specialisation independent
scheduling rules

\begin{definition}[stack-queue derivation steps]
\label{d4.1}
A set $SQ$ of derivation
steps is said to be of \emph{stack-queue} type, if it verifies the following
condition. Given any clause$\;c=(ht\longleftarrow B)$, two p-goals $M_{s}$
and $M_{q}$ exist with $M_{s}|M_{q}=B$, such that for any p-goal $(a|K)$%
:\smallskip

$a|K\overset{SQ,c\xi ,\mu }{\longrightarrow }R\;\;\;\Longrightarrow
\;\;\;R=(M_{s}\xi \underline{\gamma }|K|M_{q}\xi \underline{\gamma })\mu .$
\medskip
\end{definition}

The following property states that any set of stack-queue derivation steps
is specialisation independent. Then, as stated in Theorem \ref{t4.1}, any
set of stack-queue derivation steps which satisfies the completeness
property is a specialisation independent scheduling rule.

\begin{property}[stack-queue implies specialisation independence]
\label{py4.1}
Let $%
SQ$ be a stack-queue set of derivation steps. Then $SQ$ is specialisation
independent.
\end{property}

\begin{proof}
Let us consider two derivation steps in $SQ$ and
suppose that derivation step (2) is a lowering of (1) by $F$. This means
that (1) and (2) have the following form, where$\;c=(ht\longleftarrow
M_{s}|M_{q})$:\smallskip

$a|K\overset{SQ,c}{\longrightarrow }(M_{s}\xi ^{\prime }\underline{\gamma }%
^{\prime }|K|M_{q}\xi ^{\prime }\underline{\gamma }^{\prime })\alpha
^{\prime }$\hfill (1)\qquad \smallskip

$a\lambda \underline{\sigma }|(K\lambda \underline{\sigma }+F)\overset{SQ,c}{%
\longrightarrow }(M_{s}\xi ^{\prime \prime }\underline{\gamma }^{\prime
\prime }|(K\lambda \underline{\sigma }+F)|M_{q}\xi ^{\prime \prime }%
\underline{\gamma }^{\prime \prime })\alpha ^{\prime \prime }.$\hfill
(2)\qquad \smallskip

\noindent In order to show that $SQ$ is specialisation independent, we have
to verify that derivation step (2) is a congruent lowering of (1) by $F$,
i.e. a shifting $\underline{\rho }$ exists, such that:\smallskip

$M_{s}\underline{\gamma }^{\prime \prime }|M_{q}\underline{\gamma }^{\prime
\prime }=(M_{s}\underline{\gamma }^{\prime }|M_{q}\underline{\gamma }%
^{\prime })\underline{\rho },\;\;\;K$\underline{$\sigma $}$=K\underline{\rho 
}$.\hfill (3)\qquad \smallskip

\noindent By Property \ref{py2.2.1}, a shifting $\underline{\rho }$ exists
such that:\smallskip

$M_{s}\underline{\gamma }^{\prime \prime }|K\underline{\sigma }|M_{q}%
\underline{\gamma }^{\prime \prime }=^{(Ax-iii)}(M_{s}\underline{\gamma }%
^{\prime }|K|M_{q}\underline{\gamma }^{\prime })\underline{\rho }=M_{s}%
\underline{\gamma }^{\prime }\underline{\rho }|K\underline{\rho }|M_{q}%
\underline{\gamma }^{\prime }\underline{\rho }.$\smallskip

\noindent Since it is evident that $\#M_{s}\underline{\gamma }^{\prime }%
\underline{\rho }=\#M_{s}\underline{\gamma }^{\prime \prime }$ \ and $\#K%
\underline{\rho }=\#K\underline{\sigma }$, by Property \ref{py2.1.1}-i) we
have:\smallskip

$M_{s}\underline{\gamma }^{\prime \prime }=M_{s}\underline{\gamma }^{\prime }%
\underline{\rho },\;\;M_{q}\underline{\gamma }^{\prime \prime }=M_{q}%
\underline{\gamma }^{\prime }\underline{\rho },\;\;K\underline{\sigma }=K%
\underline{\rho }$,\smallskip

\noindent which immediately implies assertion (3).
\end{proof}

\begin{theorem}[stack-queue scheduling rules]
\label{t4.1}
Let $SQ$ be a complete
set of stack-queue derivation steps. Then $SQ$ is a specialisation
independent scheduling rule.
\end{theorem}

\subsection{\label{subs4.1}Specialisation independence implies stack-queue}

Now, we prove (Theorem \ref{t4.1.1}) that any specialisation independent
scheduling rule is actually a stack-queue rule. Thus, combining this fact
with Theorem \ref{t4.1}, we have that Definition \ref{d3.2} and the
operational characterisation of Definition \ref{d4.1} identify the same
family of scheduling rules. To this aim, let us show the following lemma.

\begin{lemma}[not internal positioning]
\label{L4.1.1}
Let$\;S\;$be a
specialisation independent scheduling rule. Given any clause$%
\;c=(ht\longleftarrow B)$, for every derivation step of the form:\smallskip

$a|K\overset{S,c\xi ,\eta }{\longrightarrow }R,$\hfill (1)\qquad 
\smallskip

\noindent two subgoals $M_{s}$ and $M_{q}$ exist, with$\;B=M_{s}|M_{q}$,
such that:\smallskip

$R=(M_{s}\xi \underline{\gamma }|K|M_{q}\xi \underline{\gamma })\eta .$
\end{lemma}

\begin{proof}
Let us consider a p-goal like:\smallskip

$a|K\underline{\omega }_{1}|K\underline{\omega }_{2}|...|K\underline{\omega }%
_{n}$, \ with $n>\#B$.\smallskip

\noindent On the basis of (1), by Property \ref{py2.5.1} a derivation step
also exists of the following form:\medskip

$a|K\underline{\omega }_{1}|...|K\underline{\omega }_{n}\overset{S,c\gamma
,\mu }{\longrightarrow }(Q=((K\underline{\omega }_{1}|...|K\underline{\omega 
}_{n})+B\gamma \underline{\tau })\mu )$.\hfill (2)\qquad \medskip

\noindent Since $n>\#B$, an index $j$ must exist such that no atom of$\;B\;$%
has been positioned inside $K\underline{\omega }_{j}$. A priori several $j$%
's might exist. Without loss of generality, we take any one of them. Thus,
two p-goals $M_{s}$ and $M_{q}$ must exist, with $M_{s}|M_{q}=B$, such
that:\medskip

$Q=(M_{s}\underline{\tau }\gamma +(K\underline{\omega }_{1}|...|K\underline{%
\omega }_{j-1}))|K\underline{\omega }_{j}|(M_{q}\underline{\tau }\gamma +(K%
\underline{\omega }_{j+1}|...|K\underline{\omega }_{n}))\mu .$\hfill
(3)\qquad \medskip

\noindent Now, by definition, derivation step (1) has the form:\smallskip

$a|K\overset{S,c}{\longrightarrow }(R=(K+B\xi \underline{\sigma })\eta )$%
.\hfill (1a)\qquad \smallskip

\noindent Since$\;S\;$is a specialisation independent rule, step (2) is a
congruent lowering of step (1a) by the subgoal $(K\underline{\omega }%
_{1}|...|K\underline{\omega }_{j-1}|K\underline{\omega }_{j+1}|...|K%
\underline{\omega }_{n})$, so that a shifting $\underline{\rho }$ exists
with $K\underline{\rho }=K\underline{\omega }_{j}$ and $B\underline{\sigma }%
\underline{\rho }=B\underline{\tau }=M_{s}\underline{\tau }|M_{q}\underline{%
\tau }$. Then, recalling that (3) implies $M_{s}\underline{\tau }\dashv K%
\underline{\omega }_{j}\dashv M_{q}\underline{\tau }$, we obtain:\medskip

$(K+B\xi \underline{\sigma })\underline{\rho }=K\underline{\omega }_{j}+B\xi 
\underline{\tau }=(M_{s}\xi \underline{\tau }|K\underline{\omega }%
_{j}|M_{q}\xi \underline{\tau })=(M_{s}\xi \underline{\tau }|K\underline{%
\rho }|M_{q}\xi \underline{\tau })$.\medskip

\noindent Finally:\medskip

$R=(K+B\xi \underline{\sigma })\eta =(K+B\xi \underline{\sigma })\eta 
\underline{\rho }\underline{\rho }^{-1}=(M_{s}\xi \underline{\tau }%
\underline{\rho }^{-1}|K|M_{q}\xi \underline{\tau }\underline{\rho }%
^{-1})\eta $.
\end{proof}

The following Theorem \ref{t4.1.1} shows that, for any scheduling rule,
specialisation independence implies that the rule is stack-queue. Together
with Theorem \ref{t4.1}, this result proofs that stack-queue is an
operational characterisation of the set of specialisation independent
scheduling rules.

\begin{theorem}[specialisation independence implies stack-queue]
\label{t4.1.1}
Let$%
\;S\;$be a specialisation independent scheduling rule. Given any clause$%
\;c=(ht\longleftarrow B),$ two p-goals $M_{s}$ and $M_{q}$ exist, with $%
M_{s}|M_{q}=B$, such that for every derivation step of the form:\smallskip

$a|K\overset{S,c\xi ,\eta }{\longrightarrow }R\hfill $(1)\qquad 
\smallskip

\noindent it is:\smallskip

$R=(M_{s}\xi \underline{\pi }|K|M_{q}\xi \underline{\pi })\eta $.
\end{theorem}

\begin{proof}
Let $p$ be the predicate symbol of atom $ht$.
Consider a p-atom $b$ of the form $b=p(x_{1},...,x_{k})[s]$, where $%
x_{1},...,x_{k}$ are distinct variables. Then, consider a ground p-atom $r$
such that $b\dashv r$. By construction of $b$ and completeness of$\;S,$ a
derivation step of the type $(b|r\overset{S,c}{\longrightarrow }\bullet )$
exists, which necessarily has the following form because $r$ is a single
atom:\smallskip

$b|r\overset{S,c}{\longrightarrow }(M_{s}\lambda \underline{\varepsilon }%
|r|M_{q}\lambda \underline{\varepsilon })\mu ,$ \ \ with$\;B=M_{s}|M_{q}$%
.\hfill (2)\qquad \smallskip

\noindent Now, let us prove that $M_{s}|M_{q}$ is the partition of$\;B\;$%
which is required by the thesis. Consider derivation step (1). Two cases are
possible, either $K=\varnothing $ or $K\neq \varnothing $.\bigskip

\noindent \underline{Case 1} $(K=\varnothing ).\medskip $

\noindent In this case we have:\smallskip

$a\overset{S,c}{\longrightarrow }(R=B\xi \underline{\pi }\eta =(M_{s}\xi 
\underline{\pi }|M_{q}\xi \underline{\pi })\eta ).$\bigskip

\noindent \underline{Case 2} $(K\neq \varnothing ).\medskip $

\noindent On the basis of (1), we have that also p-atom $a$ has $p$ as a
predicate symbol, so that a substitution $\tau $ and a shifting $\underline{%
\sigma }$ exist with $a=b\tau \underline{\sigma }$. By (1) and Property \ref
{py2.5.1}, a derivation step exists like:\smallskip

$(b\tau \underline{\sigma }|(r\tau \underline{\sigma }+K)=a|(K+r\underline{%
\sigma }))\overset{S,c\xi ^{\prime },\eta ^{\prime }}{\longrightarrow }Q$%
,\hfill (4)\qquad \smallskip

\noindent where by Lemma \ref{L4.1.1} we have that:\smallskip

$Q=(N_{s}\xi ^{\prime }\underline{\gamma }|(r\underline{\sigma }+K)|N_{q}\xi
^{\prime }\underline{\gamma })\eta ^{\prime }$, \ \ with$\;B=N_{s}|N_{q}$%
.\hfill (5)\qquad \smallskip

\noindent The proof can be now completed by exploiting derivation step (4)
as a sort of ``bridge'' between (1) and (2). In fact, since$\;S\;$is
specialisation independent rule, derivation step (4) is a congruent lowering
of step (2) by $K$, so that a shifting $\underline{\rho }^{\prime }$ exists
with $r\underline{\rho }^{\prime }=r\underline{\sigma }$ and $(M_{s}%
\underline{\varepsilon }|M_{q}\underline{\varepsilon })\underline{\rho }%
^{\prime }=N_{s}\underline{\gamma }|N_{q}\underline{\gamma }$. As a
consequence (see (5) and (2)), we can write:\medskip

$N_{s}\underline{\gamma }|r\underline{\sigma }|N_{q}\underline{\gamma }=N_{s}%
\underline{\gamma }|N_{q}\underline{\gamma }+r\underline{\sigma }=(M_{s}%
\underline{\varepsilon }|M_{q}\underline{\varepsilon })\underline{\rho }%
^{\prime }+r\underline{\rho }^{\prime }=(M_{s}\underline{\varepsilon }%
|r|M_{q}\underline{\varepsilon })\underline{\rho }^{\prime }$,\medskip

with $r\underline{\sigma }=r\underline{\rho }^{\prime }$.\smallskip

\noindent Then, by Property \ref{py2.1.1}-ii we have that $N_{s}\underline{%
\gamma }=M_{s}\underline{\varepsilon }\underline{\rho }^{\prime }$, which
obviously implies:\smallskip

$\#N_{s}=\#M_{s}$.\hfill (6)\qquad \smallskip

\noindent Now, let us note that by Lemma \ref{L4.1.1} it must be:\smallskip

$R=(A_{s}\xi \underline{\pi }|K|A_{q}\xi \underline{\pi })\eta ,$ \ \ with$%
\;B=A_{s}|A_{q}$.\hfill (7)\qquad \smallskip

\noindent Since$\;S\;$is a specialisation independent rule, derivation step
(4) is a congruent lowering of step (1) by $r\underline{\sigma }$, so that a
shifting $\underline{\rho }^{\prime \prime }$ exists with $K\underline{\rho }%
^{\prime \prime }=K$ and $(A_{s}\underline{\pi }|A_{q}\underline{\pi })%
\underline{\rho }^{\prime \prime }=N_{s}\underline{\gamma }|N_{q}\underline{%
\gamma }$. As a consequence (see (5) and (7)) we can write:\medskip

$N_{s}\underline{\gamma }|K|N_{q}\underline{\gamma }=N_{s}\underline{\gamma }%
|N_{q}\underline{\gamma }+K=(A_{s}\underline{\pi }|A_{q}\underline{\pi })%
\underline{\rho }^{\prime \prime }+K\underline{\rho }^{\prime \prime }=(A_{s}%
\underline{\pi }|K|A_{q}\underline{\pi })\underline{\rho }^{\prime \prime }$%
,\smallskip

with $K=K\underline{\rho }^{\prime \prime }$.\medskip

\noindent Then, by Property \ref{py2.1.1}-ii, we have that $N_{s}\underline{%
\gamma }=A_{s}\underline{\pi }\underline{\rho }^{\prime \prime }$, which
obviously implies:\smallskip

$\#N_{s}=\#A_{s}$.\hfill (8)\qquad \smallskip

\noindent By (2) and (7), it is $M_{s}|M_{q}=A_{s}|A_{q}=B$. By (6), (8) and
Property \ref{py2.1.1}-i), we have that:\smallskip

$A_{s}=M_{s}$ and $A_{q}=M_{q}$.\smallskip

\noindent Substituting in (7) , the thesis is obtained.
\end{proof}

\subsection{\label{subs4.2}Notes on the structure of stack-queue derivations}

Let us consider a stack-queue derivation like:\smallskip

$A|B\overset{SQ,M,\sigma }{\longrightarrow }\bullet $, \ \ where $%
M=c_{1},c_{2},...c_{h}$ and $M/B=\varnothing .\hfill $(1)\qquad \smallskip

\noindent By definition of stack-queue scheduling rules, only atoms in $A$
together with atoms deriving from $A$ and allocated in stack mode can be
rewritten in derivation (1). Thus, derivation (1) has the form:\smallskip

\noindent $A|B\overset{SQ,c_{1}\xi _{1},\sigma _{1}}{\longrightarrow }%
X_{1}|A_{1}|B\sigma _{1}|Y_{1}\overset{SQ,c_{2}\xi _{2},\sigma _{2}}{%
\longrightarrow }...$\smallskip

$X_{i}|A_{i}|B\sigma _{1}...\sigma _{i}|Y_{i}\overset{SQ,c_{i+1}\xi
_{i+1},\sigma _{i+1}}{\longrightarrow }...\overset{SQ,c_{h}\xi _{h},\sigma
_{h}}{\longrightarrow }X_{h}|A_{h}|B\sigma _{1}...\sigma _{h}|Y_{h}$,\hfill
(1a)\qquad \smallskip

\noindent where:

\begin{itemize}
\item  each $X_{i}$ is formed by new atoms deriving from $A$ which are
allocated in stack mode,

\item  each $A_{i}$ is formed by atoms of $A$ which are not yet rewritten,

\item  each $Y_{i}$ is formed by new atoms deriving from $A$ which are
allocated in queue mode.
\end{itemize}

\noindent The above structural considerations suggest the following formal
definition.

\begin{definition}[A-preq type derivations]
\label{d4.2.1}
A p-SLD
derivation, of the form $A|B\overset{SQ,M}{\longrightarrow }\bullet $, is of%
\emph{\ pre-queued type w.r.t.} the subgoal $A$ (simply written $A$\emph{%
-preq type} in the following) if the only rewritten atoms are:\smallskip

- atoms from the subgoal $A$,

- atoms deriving from $A$ and allocated in stack mode.\medskip
\end{definition}

Note that Definition \ref{d4.2.1} is significant even if $B=\varnothing $.
It is evident that any $A$-preq derivation has the form (1a). In the sequel
we use the following shortened notation to represent $A$-preq type
derivations:\smallskip

$A|B\overset{SQ,M,\sigma }{\longrightarrow }A^{s}|B\sigma |A^{q},$\hfill
(Ap)\qquad \smallskip

\noindent where, with reference to (1a), $A^{s}=X_{h}|A_{h}$ stands for
``stacked subgoal derived from $A$'', and $A^{q}=Y_{h}$ means ``queued
subgoal derived from $A$''. It is evident that in any preq type derivation
we have $M/B=\varnothing $.

The following definition characterises an $A$\emph{-queued} derivation as an 
$A$-preq derivation where all atoms of $A$ are rewritten together with all
atoms deriving from $A$ and allocated in stack mode, i.e. $A^{s}$= $%
\varnothing $. Intuitively, an $A$-queued derivation is an $A$-preq
derivation which cannot be extended without loosing its $A$-preq nature.
Indeed, the acronym ``$A$-preq'' stands for ``$A$-pre-queued'' derivation.

\begin{definition}[A-queued derivations]
\label{d4.2.2}
Let $SQ$ be a
stack-queue scheduling rule. A derivation which is of $A$-preq type and has
the form:\smallskip

$A|B\overset{SQ,K,\sigma }{\longrightarrow }B\sigma |A^{q}$\hfill \emph{%
(Aq)\qquad }\smallskip

\noindent is said to be \emph{queued w.r.t. }$A$ (simply written $A$\emph{%
-queued} in the following).\medskip
\end{definition}

In the following Section \ref{subs4.3}, we will exploit the notations
introduced in (Ap) and (Aq) to represent $A$-preq type and $A$-queued
derivations, respectively. It is worth noting that starting from a p-goal of
the form $A|B$, when the $A$-queued derivation is reached, the last
resolvent presents a situation where the roles of $A$ and$\;B\;$are
exchanged. In practice, restarting from $B\sigma |A^{q}$, the derivation can
attempt to proceed towards a $(B\sigma )$-queued derivation. The proof of an
important result in Section \ref{subs4.3} (Duplication Theorem \ref{t4.3.1})
is based on this cyclic behaviour of stack-queue derivations.

\subsection{\label{subs4.3}Duplication tolerance}

In this section an important property is shown for stack-queue scheduling
rules. Let us give an intuitive presentation of this result, which is stated
in the \emph{full duplication theorem} (Theorem \ref{t4.3.2}). Suppose that
a p-SLD derivation $Dr$ of$\;G\;$in P can be developed via a stack-queue
scheduling rule $SQ$. Then consider a p-goal $G^{\prime }$ which is equal to$%
\;G\;$apart from the duplication of some atoms. Furthermore, suppose that
each copy is scheduled after the corresponding original atom. In this
hypothesis, the full duplication theorem asserts that a p-SLD derivation of $%
G^{\prime }$ in P exists via the same scheduling rule $SQ$, where all
derivation steps of $Dr$ are redone in the order.

The full duplication theorem is basic for the proof of the final results of
the paper, i.e. results about redundancy elimination tolerance which are
given in Section \ref{s5.reduElim}. Indeed, let us consider the problem of
preserving program termination. Intuitively, program termination is
preserved if the introduction of redundancy elimination does not provoke any
really different new derivations. Reversing the viewpoint, termination is
retained if any derivation, developed in presence of redundancy elimination,
can be traced again when redundancy is left in place. The full duplication
theorem asserts this kind of fact in the simplest case, i.e. when redundancy
has the form of a replica of atoms already present in the initial p-goal,
provided that the scheduling rule is of stack-queue type.

First we show a duplication theorem (Theorem \ref{t4.3.1}) which is valid
when only one atom or group of adjacent atoms is duplicated. Then the result
is easily extended to obtain the full theorem. Though intuitive in
appearance, Theorem \ref{t4.3.1} has a relatively complex proof. In this
section we give only a sketch of the argument. In the sketch, we will make
reference to the particular case of completely ground derivations, i.e.
derivations such that all resolvents are ground. This simplification will
allow us to highlight the essence of the argument, without having to do with
technical problems deriving from variable instantiations. Formal
presentation of the proof of Theorem \ref{t4.3.1} is given in Appendix B.
Note that the hypothesis of ground resolvents is verified in the case that
no new variable is present in clause bodies and initial goals are ground.
\medskip

\begin{theorem}[duplication theorem]
\label{t4.3.1}
Let P be a logic program and $SQ$
a stack-queue scheduling rule. Given two p-goals of the form $A|B|C|D$ and $%
A|B|C|B\underline{\pi }|D$, the following implication holds:\smallskip

$A|B|C|D\overset{SQ,X.P}{\longrightarrow }Q$\hfill (1)\qquad 
\smallskip

$\qquad \Longrightarrow \;\exists Y$ such that \ $A|B|C|B\underline{\pi }|D%
\overset{SQ,Y.P}{\longrightarrow }R$\smallskip

\qquad \qquad\ \ \ with $X\subseteq _{L}Y$ \ \ and \ $\#Q\preceq \#R.$
\smallskip
\end{theorem}

\begin{proof}[Proof (sketch)]
Let $\Delta (SQ,n)$ denote the
subset of $\Delta (SQ)$ such that, for any derivation $Dr$ in $\Delta (SQ,n)$
, it is $\#Dr\preceq n$, where $\#Dr$ denotes the length of $Dr$. We show
the thesis by induction on $n$. In other words, we show that the thesis
holds when derivation (1) belongs to $\Delta (SQ,n)$, for any $n\succeq 0$.
The fact is obvious for $\Delta (SQ,0)$. In order to justify the inductive
step from $\Delta (SQ,n-1)$ to $\Delta (SQ,n)$, for $n>0$, let us consider a
derivation like:\smallskip

$(A|B|C|D\overset{X.P}{\longrightarrow }Q)\in \Delta (SQ,n)\hfill $%
(1a)\qquad \smallskip

\noindent and show that $(A|B|C|B\underline{\pi }|D\overset{SQ,Y.P}{%
\longrightarrow }R)$ exists with $X\subseteq _{L}Y$ and $\#Q\preceq \#R$.
The following three possible situations must be taken into account. Then, we
start with case 3, which is the most significant one.

\begin{enumerate}
\item  derivation (1a) is of $(A|B|C)$-preq type,\smallskip

\item  derivation (1a) is of $(A|B|C|D)$-preq type, and not of $(A|B|C)$
-preq type,\smallskip

\item  derivation (1a) is not of $(A|B|C|D)$-preq type.\smallskip
\end{enumerate}

\noindent \underline{Case 3}.\smallskip

\noindent As already said, the simplified argument, which we use in this
sketch, works in the hypothesis that all resolvents are ground, so that
derivation (1a) has the following form:\smallskip

\noindent $A|B|C|D\overset{H}{\longrightarrow }B|C|D|A^{q}\overset{K}{%
\longrightarrow }C|D|A^{q}|B^{q}\overset{M}{\longrightarrow }$\smallskip

$D|A^{q}|B^{q}|C^{q}\overset{N}{\longrightarrow }A^{q}|B^{q}|C^{q}|D^{q}%
\overset{T}{\longrightarrow }Q$, where $H|K|M|N|T=X$.\hfill (2)\qquad
\smallskip

\noindent Then, it is intuitive that a derivation can be constructed like
the following, where $\underline{\phi }$ is a suitable shifting:\smallskip

\noindent $A|B|C|B\underline{\pi }|D\overset{SQ,H}{\longrightarrow }B|C|B%
\underline{\pi }|D|A^{q}\overset{SQ,K}{\longrightarrow }$\smallskip

$C|B\underline{\pi }|D|A^{q}|B^{q}\overset{SQ,M}{\longrightarrow }B%
\underline{\pi }|D|A^{q}|B^{q}|C^{q}\overset{SQ,K}{\longrightarrow }$\hfill
(3)\qquad \smallskip

$D|A^{q}|B^{q}|C^{q}|B^{q}\underline{\phi }\overset{SQ,N}{\longrightarrow }%
A^{q}|B^{q}|C^{q}|B^{q}\underline{\phi }|D^{q}.$\smallskip

\noindent By construction of (2), $A^{q}|B^{q}|C^{q}|D^{q}\overset{T}{%
\longrightarrow }Q$ is a derivation belonging to $\Delta (SQ,m)$, with $m<n$%
. By inductive hypothesis, a derivation exists such that:

$A^{q}|B^{q}|C^{q}|B^{q}\underline{\phi }|D^{q}\overset{SQ,Y^{\prime }.P}{%
\longrightarrow }R^{\prime }$,\hfill (5)\qquad \smallskip

with $T\subseteq _{L}Y^{\prime }$ and $\#Q\preceq \#R^{\prime }$.\hfill
(5a)\qquad \smallskip

\noindent By Property \ref{py3.2.1}, derivations (3) and (5) can be combined
to yield a derivation of the form:

$A|B|C|B\underline{\pi }|D\overset{SQ,(H|K|M|K|N).P}{\longrightarrow }
A^{q}|B^{q}|C^{q}|B^{q}\underline{\phi }|D^{q}\overset{SQ,Y^{\prime }.P}{%
\longrightarrow }R$,\hfill (6)\qquad \smallskip

\noindent where$\;R\;$is a p-variant of $R^{\prime }$, which implies $
\#R=\#R^{\prime }$. Finally:\smallskip

$X=H|K|M|N|T\subseteq _{L}^{(5a)}(H|K|M|K|N|Y^{\prime }),\;\;\#Q\preceq
^{(5a)}\#R^{\prime }=\#R$.\bigskip

\noindent \underline{Case 2}.\smallskip

\noindent Derivation (1a) has the form $A|B|C|D\overset{H|K|M|N}{%
\longrightarrow }D^{s}|A^{q}|B^{q}|C^{q}|D^{q}$, where $H|K|M|N=X$.
Analogously to case 3), a derivation can be constructed like:\smallskip

$A|B|C|B\underline{\pi }|D\overset{SQ,(H|K|M|K|N).P}{\longrightarrow }%
D^{s}|A^{q}|B^{q}|C^{q}|B^{q}\underline{\pi }|D^{q}$.\bigskip

\noindent \underline{Case 1}.\smallskip

\noindent Derivation (1a) has the form $A|B|C|D\overset{X}{\longrightarrow }%
(A|B|C)^{s}|D|(A|B|C)^{q}$. A derivation exists like:\smallskip

$(A|B|C)|B\underline{\pi }|D\overset{SQ,X}{\longrightarrow }(A|B|C)^{s}|B
\underline{\pi }|D|(A|B|C)^{q}.$
\end{proof}

Now we can state and prove the full duplication theorem, which extends the
previous Theorem \ref{t4.3.1} to the duplication of two or more not adjacent
atoms in the initial goal of a p-SLD derivation.
\bigskip

\begin{theorem}[full duplication theorem]
\label{t4.3.2}
Let P be a logic program
and $SQ$ a stack-queue scheduling rule. Given a p-goal $N+F$ such
that:\smallskip

$\forall b[s]\in F,\;\;\;\exists b[s^{\prime }]\in N$\ \ \ with $s^{\prime
}<s$,\smallskip

\noindent the following implication holds:\smallskip

$N\overset{SQ,M.P}{\longrightarrow }Q$\smallskip

$\qquad \Longrightarrow $ $\exists Y$ such that $N+F\overset{SQ,Y.P}{%
\longrightarrow }R$\smallskip

\qquad \qquad\ \ with $M\subseteq _{L}Y$ and $\#Q\preceq \#R$.
\end{theorem}

\begin{proof}
By hypothesis, the subgoal$\;F\;$is made of
duplicated atoms. Then, the proof is by induction on the length of $F$.
Indeed, if$\;F\;$is empty the thesis is true. Now, suppose that the thesis
is already proven for any$\;F\;$with $\#F=n\succeq 0$. Then let us consider
any p-goal$\;G=F|b[s]$ with $\#F=n$. By inductive hypothesis a derivation
exists such that:\smallskip

$N+F\overset{SQ,Z.P}{\longrightarrow }S,$ \ \ with $\ \ M\subseteq _{L}Z$ \
\ and $\ \ \#Q\preceq \#S$.\smallskip

\noindent By hypothesis, three p-goals $A,C$ and $D$ exist together with a
p-atom $b[s^{\prime }]$, such that:\smallskip

$N+(F|b[s])=A|b[s^{\prime }]|C|b[s]|D$ \ \ and \ $N+F=A|b[s^{\prime }]|C|D$%
.\smallskip

\noindent As a consequence, Theorem \ref{t4.3.1} can be applied to $N+F$ and 
$N+(F|b[s])$ yielding:\smallskip

$N+(F|b[s])\overset{SQ,Y.P}{\longrightarrow }R$, \ \ with $Z\subseteq _{L}Y$
and $\#S\preceq \#R$.\smallskip

\noindent Now the induction step is completed, because:\smallskip

$M\subseteq _{L}Z\subseteq _{L}Y$ \ \ and \ \ $\#Q\preceq \#S\preceq \#R$.
\end{proof}

\section{\label{s5.reduElim}Redundancy elimination tolerance}

In this section, the tolerance of stack-queue scheduling rules to redundancy
elimination is considered. The preservation of program termination in shown
in Section \ref{subs5.1}. The preservation of the completeness of $EVR_{L}$
loop check is shown in Section \ref{subs5.2} for function free programs.
First, the idea of goal reduction, which is originally given in \cite
{6.FPS95} and is recalled in Definition \ref{d1.1} of this paper, is
restated. Indeed, in Section \ref{s1.goalRedu} little attention is paid to
the positions of atoms which are removed from a resolvent. However, if the
execution is based on atom priority values, it is intuitive that removing an
atom without any convenient expedient may overthrow the essence of previous
atom scheduling. Thus, a refined definition of goal reduction is given below
(Definition \ref{d5.1}) which fits the frame of priority SLD derivation
mechanisms.

The inspiring idea of \emph{priority reduction} is quite simple. According
to Definition \ref{d1.1}, for any removed atom $b$, an \emph{eliminating}
atom $a=b\tau $ exists which remains in the reduced resolvent. Several
removed atoms may share the same eliminating one. In reference to Definition 
\ref{d5.1} below, for any eliminating atom $a_{j}[p_{j}]$, the corresponding
subset $A_{j}$ of eliminated atoms is pointed out. Then, except for the case 
$a_{j}[p_{j}]\dashv A_{j}$, any $a_{j}[p_{j}]$ is advanced to the least
priority value in $A_{j}$. In other words, \emph{each eliminating atom is
advanced to replace the first scheduled atom among its eliminated ones}.
Intuitively, the aim is to restore the essence of the previous atom
priorities. The notation $\{+A_{j},1\preceq j\preceq h\}$ will represent the
merging $A_{1}+A_{2}+...+A_{h}$, and the notation $prs(A_{j})$ the set of
priority values in $A_{j}$.
\bigskip

\begin{definition}[priority reduced goals]
\label{d5.1}
Let $X$ be a set of variables, $%
\tau $ a substitution and$\;G\;$a p-goal. A p-goal$\;N\;$is a \emph{reduced
p-goal} of$\;G\;$by $\tau $ up to X, denoted by$\;G>>^{\tau }N$, if the
following conditions hold:\medskip

i)$\;G\;=\;F\;+\;\{+a_{j}[p_{j}],\;1\preceq j\preceq
h\}\;+\;\{+A_{j},\;1\preceq j\preceq h\}$,\smallskip

\qquad where $\forall b[s]\in A_{j},\;\;\;b\tau =a_{j},\;\;\;1\preceq
j\preceq h$,\medskip

ii)$\;N\;=\;F\;+\;\{+a_{j}[r_{j}],\;1\preceq j\preceq h\},$\smallskip

\qquad where $r_{j}=min(\{p_{j}\}\cup prs(A_{j})),\;\;\;1\preceq j\preceq h,$%
\medskip

iii) $\forall x\in (X\cup var(N))$ \ \ it is \ $x\tau =x$.\medskip
\end{definition}

\begin{example}
\noindent Given the p-goal\smallskip

$G=p(z)[1],q(w)[2],p(a)[3],p(y)[4],q(v)[5]$,\smallskip

\noindent the following$\;N\;$is a reduced p-goal of$\;G\;$by the
substitution $\tau =\{z/a,y/a,v/w\}$:\smallskip

$N=p(a)[1],q(w)[2]$.\smallskip

\noindent Note that $p(a)[3]$ has been advanced to replace the first of the
atoms it eliminates, that is $p(z)[1]$.~$\square$
\end{example}
\medskip

Now, the idea of \emph{priority reduced SLD derivation} can be defined as a
generalisation of Definition \ref{d2.3.2}. In essence a priority reduced SLD
derivation is a p-SLD derivation where, at any step, a priority reduction of
the resolvent according to Definition \ref{d5.1} is allowed.
\bigskip

\begin{definition}[priority Reduced SLD derivation]
\label{d5.2} 
Let P be a program and 
$G_{o}$ a p-goal. A \emph{priority reduced SLD derivation} of $G_{o}$ in P (%
\emph{p-RSLD derivation }for short) is a possibly infinite sequence of
priority reductions and derivation steps\smallskip

$G_{o}>>^{\alpha _{o}}N_{o}\overset{c_{o}\xi _{o},\theta _{o}}{
\longrightarrow }G_{1}\;...\;G_{k}>>^{\alpha _{k}}N_{k}\overset{c_{k}\xi
_{k},\theta _{k}}{\longrightarrow }G_{k+1}>>^{\alpha _{k+1}}N_{k+1}\;...$
\smallskip

\noindent where, for any $j\succeq 0$,\smallskip

i) $c_{j}$ is a clause in P,\smallskip

ii) $var(c_{j}\xi _{j})\cap (var(G_{o})\cup var(c_{o}\xi _{o})\cup ...\cup
var(c_{j-1}\xi _{j-1}))=\varnothing $,\smallskip

iii) $G_{j}>>^{\alpha _{j}}N_{j}$ \ up to $var(G_{o}\theta _{o}...\theta
_{j-1})$.\medskip
\end{definition}

\noindent The notation\smallskip

$G\overset{S,D}{\longrightarrow }>>N$\smallskip

\noindent will be used to represent a p-RSLD derivation which is developed
in agreement with the scheduling rule$\;S\;$using the template $D$. The last
resolvent$\;N\;$is intended to be a reduced resolvent.

\subsection{\label{subs5.1}Termination preserving}

In this section, the redundancy elimination tolerance of stack-queue
scheduling rules is shown, with reference to program termination (Theorem 
\ref{t5.1.1}). The following lemma is fundamental for proving the
preservation of termination, as well as the preservation of $EVR_{L}$ loop
check completeness.
\medskip

\begin{lemma}
\label{L5.1.1}Let P be a program and $SQ$ a stack-queue scheduling rule. The
following implication holds:\smallskip

$G\overset{SQ,X.P}{\longrightarrow }>>Q\hfill $(1)\qquad \smallskip

$\qquad \Longrightarrow \;\;\exists Z$\ \ such that$\;G\overset{SQ,Z.P}{%
\longrightarrow }R$, \ \ \ with $X\subseteq _{L}Z,\;\#Q\preceq \#R$.
\end{lemma}

\begin{proof}
\noindent The proof is by induction on the length of $X$. If $\#X=0$, the
thesis is trivially verified with $Z=\varnothing $. Then let us consider $%
X=c|H$. Derivation (1) may be rewritten as:\smallskip

$(G>>^{\tau }N)\overset{SQ,c}{\longrightarrow }F\overset{SQ,H}{%
\longrightarrow }>>Q.$\hfill (2)\qquad \smallskip

\noindent Since $\#H<\#X$, by inductive hypothesis, a p-SLD derivation
exists of the form:\smallskip

$F\overset{SQ,K.P}{\longrightarrow }T$, \ \ with $H\subseteq
_{L}K,\;\#Q\preceq \#T$.\hfill (3)\qquad \smallskip

\noindent By Property \ref{py3.2.1}, the first derivation step of (2) and
derivation (3) can be combined to yield a derivation of the following
form:\smallskip

$N\overset{SQ,c}{\longrightarrow }F\overset{SQ,K}{\longrightarrow }S,$ \ \
where$\;S\;$is a p-variant of $T$.\hfill (4)\qquad \smallskip

\noindent Now, let us consider the p-goal $G\tau $. With reference to
Definition \ref{d5.1}, we have that:\smallskip

\noindent $G\tau =(F+\{+a_{j}[p_{j}],\;1\preceq j\preceq h\})\tau
+\{+A_{j}\tau ,\;1\preceq j\preceq h\}=^{(Def.\;\ref{d5.1}-iii)}$\smallskip

$F+\{+a_{j}[p_{j}],\;1\preceq j\preceq h\}+\{+A_{j}\tau ,\;1\preceq j\preceq
h\}=^{(Def.\;\ref{d5.1}-i-ii)}$\smallskip

$F+\{+a_{j}[r_{j}],\;1\preceq j\preceq h\}+\{+A_{j}\tau
\{r_{j}/p_{j}\},\;1\preceq j\preceq h\}=$\smallskip

$\hfill =N+\{+A_{j}\tau \{r_{j}/p_{j}\},\;1\preceq j\preceq h\}^{[}$%
\footnote{%
The notation $A_{j}\tau \{r_{j}/p_{j}\}$ means that the priority value $%
r_{j} $ is replaced by $p_{j}$ in the p-goal $A_{j}\tau $.}$^{]}$,\qquad
\smallskip

\noindent where $a_{j}[r_{j}]\dashv A_{j}\tau \{r_{j}/p_{j}\}$ and any atom
in $A_{j}\tau \{r_{j}/p_{j}\}$ is a duplicate of $a_{j},$ $1\preceq j\preceq
h$. Then,$\;N\;$and $G\tau $ verify the hypothesis of Theorem \ref{t4.3.2}.
As a consequence, by (4) a derivation also exists such that:\smallskip

$G\tau \overset{SQ,Z.P}{\longrightarrow }V,$ \ \ with $c|K\subseteq _{L}Z$
and $\#S\preceq \#V$\hfill (5)\qquad \smallskip

\noindent Now, let us apply Lifting Lemma \ref{L3.2.1} to (5). We obtain
that a p-SLD derivation exists like:\smallskip

$G\overset{SQ,Z.P}{\longrightarrow }R$.\hfill (6)\qquad \smallskip

\noindent where, applying Lowering Lemma \ref{L3.1.1} to (5) and (6), we
have that $\#V=\#R$. Finally, we conclude:\smallskip

$X=c|H\subseteq _{L}^{(3)}c|K\subseteq _{L}^{(5)}Z,$\smallskip

$\#Q\preceq ^{(3)}\#T=^{(4)}\#S\preceq ^{(5)}\#V=\#R.$
\end{proof}
\medskip

\begin{theorem}[termination preserving]
\label{t5.1.1}
Let P be a program,$\;G\;$a
p-goal and $SQ$ a stack-queue scheduling rule. If every p-SLD derivation of$%
\;G\;$in P via $SQ$ is finite, then any p-RSLD derivation via $SQ$ is finite
too.
\end{theorem}

\begin{proof}
Let $T$ be the p-SLD tree of$\;G\;$in P via $SQ$. By
hypothesis, every p-SLD derivation of$\;G\;$in P via $SQ$ is finite. As a
consequence, since $T$ is a finitely branching tree, by Konig's lemma (see
Theorem K, in \cite{9.K97}) $T$ is a finite tree. Let $f$ be the depth of $T$%
. Given any p-RSLD derivation of the form$\;G\overset{SQ,X.P}{%
\longrightarrow }>>\bullet $, by Lemma \ref{L5.1.1} a p-SLD derivation of
the form$\;G\overset{SQ,Z.P}{\longrightarrow }\bullet $ exists in $T$, with $%
X\subseteq _{L}Z$. But $\#Z\preceq f$, so that we obtain $\#X\preceq
\#Z\preceq f.$ In conclusion, the length of all p-RSLD derivations of$\;G\;$%
in P via $SQ$ is limited by $f$.
\end{proof}
\bigskip

Let us close this section with two examples which show that both stack-queue
scheduling and eliminating atom advancement are essential for redundancy
elimination tolerance. The first example shows the necessity of advancement
of eliminating atoms. The second one is an example of state scheduling rule
which is not tolerant to redundancy elimination, though goal reduction is
performed in agreement with Definition \ref{d5.1}. Of course, the scheduling
rule is not of stack-queue type. In the following sketches of p-SLD and
p-RSLD derivations, explicit indication of priority values is omitted, for
the sake of brevity.
\bigskip

\begin{example}
\label{e5.1.1} 

\noindent Let us consider the stack scheduling rule (i.e. the usual leftmost
rule) and the following single clause program P:\smallskip

$c=p\longleftarrow q(x)|p$.\smallskip

\noindent It is evident that all p-SLD derivations fail. However, if
advancement of eliminating atoms is not performed, an infinite p-RSLD
derivation of P exists, as shown in Figure \ref{f3}. $\square$

\begin{figure}[h]
\raggedright \qquad \qquad \underline{Resolvent}\qquad \qquad \qquad 
\qquad \qquad \qquad\ \qquad \underline{Reduced Resolvents}\smallskip \\

\qquad \qquad $p|q(a)>>^{\varepsilon }\qquad \qquad \qquad \qquad \qquad 
\qquad \ \ p|q(a)\overset{S,c}{\longrightarrow }$\smallskip \\

\qquad \qquad $q(x_{1})|p|q(a)>>^{\{x_{1}/a\}}\qquad \qquad \qquad 
\qquad p|q(a)\overset{S,c}{\longrightarrow }$\smallskip

\qquad \qquad $q(x_{2})|p|q(a)>>^{\{x_{2}/a\}}\qquad \qquad \qquad 
\qquad p|q(a)\overset{S,c}{\longrightarrow }$\smallskip

\qquad \qquad ......................................
\caption{} \label{f3}
\end{figure}

\end{example}

\begin{example}
\label{e5.1.2}

\noindent Let$\;S\;$be a scheduling rule which behaves as a stack rule, with
an exception when atoms having $s$ as a predicate symbol are rewritten. In
this case new atoms are positioned immediately after the first old atom, if
one exists. Then, let us consider the logic program P consisting of the
following clauses:\smallskip

$c1=r\longleftarrow $\smallskip

$c2=s(x,y)\longleftarrow t(x,y)$\smallskip

$c3=q(x,y)\longleftarrow r|s(z,y)|r|q(x,z)$.\smallskip

\noindent It is easy to verify that all p-SLD derivations of P terminate
independently of the initial p-goal. In fact, given a p-SLD derivation of$%
\;G\;$in P, where$\;G\;$is any p-goal, two cases are possible: either an
atom with predicate symbol $q$ is rewritten or not. If no atom with
predicate symbol $q$ is rewritten, the derivation terminates evidently.
Otherwise the derivation fails, as described below:\smallskip

$G\overset{S}{\longrightarrow }q(..)|K\overset{S,c3}{\longrightarrow }%
r|s(..)|r|q(..)|K\overset{S,c1}{\longrightarrow }$\smallskip

$\qquad s(..)|r|q(..)|K\overset{S,c2}{\longrightarrow }r|t(..)|q(..)|K%
\overset{S,c1}{\longrightarrow }t(..)|q(..)|K.$\smallskip

\noindent Now let us show that, if reduction of resolvents is allowed, an
infinite p-RSLD derivation of P exists.

\begin{figure}[h]
\raggedright \underline{Resolvents}\qquad \qquad \qquad 
\qquad \qquad \qquad\ \ \ \underline{Reduced Resolvents}\smallskip \\

$q(x,x_{1})|t(x_{1},x)>>^{\varepsilon }\qquad \qquad \qquad \qquad
q(x,x_{1})|t(x_{1},x)\overset{S,c3}{\longrightarrow }$\smallskip \\

$r|s(x_{2},x_{1})|r|q(x,x_{2})|t(x_{1},x)>>^{\varepsilon }\qquad \
r|s(x_{2},x_{1})|q(x,x_{2})|t(x_{1},x)\overset{S,c1}
{\longrightarrow }$\smallskip \\

$s(x_{2},x_{1})|q(x,x_{2})|t(x_{1},x)>>^{\varepsilon }\qquad \qquad
s(x_{2},x_{1})|q(x,x_{2})|t(x_{1},x)\overset{S,c2}
{\longrightarrow }$\smallskip \\

$q(x,x_{2})|t(x_{2},x_{1})|t(x_{1},x)>>^{\varepsilon }\qquad \qquad
q(x,x_{2})|t(x_{2},x_{1})|t(x_{1},x)\overset{S,c3}
{\longrightarrow }$\smallskip \\

...................................... \\
\caption{} \label{f4}
\end{figure}

\noindent It is easy to verify that the infinite RSLD derivation in Figure \ref{f4}
cannot be pruned neither by $EVR_{L}$ loop check nor by more powerful checks
(like $SIR_{M}$) which are based on \emph{subsumption} relationships between
resultants \cite{4.BAK91}.~$\square$
\end{example}

\subsection{\label{subs5.2}Preserving the completeness of $EVR_{L}$ loop
check}

In this section we prove the preservation of $EVR_{L}$ loop check
completeness, passing from p-SLD to p-RSLD. The result holds for function
free programs, provided that stack-queue scheduling rules are used in
combination with priority reduction of resolvents, as introduced in
Definition \ref{d5.1}. The section starts with a characterisation of $%
EVR_{L} $ loop check which exploits the concept of priority shifting and is
equivalent to the one stated in Definition \ref{d1.2.1}. In essence, passing
from Definition \ref{d1.2.1} to Definition \ref{d5.2.1} below, only
assertion ii) is modified. On the other hand, the requirement $%
N_{j}=N_{i}\tau \underline{\tau }$ is plainly equivalent to $N_{i}\tau
=_{L}N_{j}$, since any shifting $\underline{\tau }$ implies that the order
of atoms is preserved.
\bigskip

\begin{definition}[priority Equality Variant Check for Resultants]
\label{d5.2.1} 
A p-RSLD derivation\smallskip

$G_{o}>>^{\alpha _{o}}N_{o}\overset{c_{o}\xi _{o},\theta _{o}}{%
\longrightarrow }G_{1}\;...\;G_{h-1}>>^{\alpha _{h-1}}N_{h-1}\overset{%
c_{h-1}\xi _{h-1},\theta _{h-1}}{\longrightarrow }G_{h}>>^{\alpha
_{h}}N_{h}...$\smallskip

\noindent is \emph{pruned} by \emph{priority Equality Variant of Resultant}
check (called \emph{p-}$EVR_{L}$\emph{\ check}, in the following), if for
some $i$ and $j$, with $0\preceq i<j$, a renaming $\tau $ and a shifting $%
\underline{\tau }$ exist such that:\smallskip

i) $G_{o}\theta _{o}...\theta _{j-1}=G_{o}\theta _{o}...\theta _{i-1}\tau $%
,\smallskip

ii) $N_{j}=N_{i}\tau \underline{\tau }$.\medskip
\end{definition}

With reference to the above definition, any couple $Rs_{h}=[N_{h},G_{o}%
\theta _{o}...\theta _{h-1}]$ is a \emph{reduced resultant}. Given two
reduced resultants $Rs_{j}=[N_{j},G_{o}\theta _{o}...\theta _{j-1}]$ and $%
Rs_{i}=[N_{i},G_{o}\theta _{o}...\theta _{i-1}]$, for which requirements i)
and ii) of Definition \ref{d5.2.1} hold, we will write $Rs_{i}\cong Rs_{j}$.
In other words, Definition \ref{d5.2.1} expresses that p-$EVR_{L}$ loop
check is based on detecting that a reduced resultant is obtained which is
connected by the relationship $\cong $ to a preceding one in the same
derivation. It is worth noting that $\cong $ is an equivalence relationship.

Now let us prove Theorem \ref{t5.2.1}, which states that the completeness of
p-$EVR_{L}$ loop check is preserved passing from p-SLD to p-RSLD, if
stack-queue scheduling rules are used. To this aim we provide a necessary
condition which holds whenever p-$EVR_{L}$ prunes every infinite p-SLD
derivation of a goal$\;G\;$in a program P via a scheduling rule$\;S.$
Indeed, as shown in Lemma \ref{L5.2.1}, in this hypothesis the length of
resolvents of all possible derivations of$\;G\;$in P via$\;S\;$is limited.
The structure and the proof of Lemma \ref{L5.2.1} are strictly analogous to
the ones of Lemma \ref{L1.2.1}. Note also that Lemma \ref{L5.2.1} holds for
any scheduling rule. On the contrary, the stack-queue hypothesis is
necessary in Theorem \ref{t5.2.1}, which concludes the section.
\bigskip

\begin{lemma}
\label{L5.2.1}Let P be a program and$\;G\;$a p-goal. Suppose that all
infinite p-SLD derivations of$\;G\;$in P via a scheduling rule$\;S\;$are
pruned by p-$EVR_{L}$. Then, a finite bound $l$ exists such that, for each
resolvent$\;R\;$in any p-SLD derivation of$\;G\;$in P via$\;S,$ it is $%
\#R\preceq l$.
\end{lemma}

\begin{proof}
The proof of this lemma can be obtained from the
one of Lemma \ref{L1.2.1}, by means of the following replacements:

\noindent ``Let $T$ be the p-SLD tree of$\;G\;$in P via $S$'' for ``Let $T$
be an S-tree of$\;G\;$in P'',

\noindent ``By Determinism Lemma \ref{L3.2.2}'' for ``Since $T$ contains all
SLD derivations of$\;G\;$in P'',

\noindent ``p-$EVR_{L}$'' and ``p-variant'' for ``$EVR_{L}$'' and
``variant'', respectively.
\end{proof}
\bigskip

\begin{theorem}[p-$EVR_{L}$ loop check completeness preservation]
\label{t5.2.1}
Let P be a function free program, $G_{o}$ a p-goal and $SQ$ a stack-queue
scheduling rule. Suppose that all infinite p-SLD derivations of $G_{o}$ in P
via $SQ$ are pruned by p-$EVR_{L}$, then all infinite p-RSLD derivations of $%
G_{o}$ in P via $SQ$ are pruned by p-$EVR_{L}$.
\end{theorem}

\begin{proof}
Let $D$ be an infinite p-RSLD derivation of $G_{o}$
in P via $SQ$. Let $(G_{o}\overset{SQ,X}{\longrightarrow }>>Q)$ be any
finite prefix of $D$. By Lemma \ref{L5.1.1}, a p-SLD derivation $D^{\prime
}=(G_{o}\overset{SQ,Z}{\longrightarrow }R)$ exists with $\#Q\preceq \#R$. On
the other hand, by Lemma \ref{L5.2.1} a bound $l$ exists such that $%
\#Q\preceq \#R\preceq l$. But $Q$ is the generic reduced resolvent in $D$,
so that the number of atoms in all reduced resolvents of $D$ is bounded by $%
l $. As a consequence, the number of atoms in all reduced resultants of $D$
is also limited. Since the program P has finite many predicate symbols and
constants and no function symbol is allowed, the relationship $\cong $
between reduced resultants of $D$ has only finitely many equivalence
classes. Then, for some $0\preceq i<k$ in $D$, we have that the $k^{th}$
reduced resultant is related by $\cong $ to the $i^{th}$ one. This implies
that $D$ is pruned by p-$EVR_{L}$.
\end{proof}

\section{Conclusions}

In the paper the problem of possible undesirable effects of redundancy
elimination from resolvents is addressed. In particular we have shown that
program termination and loop check completeness can be lost. Conditions are
characterised which ensure the redundancy elimination tolerance, in the
sense that program termination and completeness of equality loop check are
preserved when redundancy is eliminated. However, difficulties in analysing
interdependence of redundancy elimination effects from the used selection
rule have arisen, and the necessity of a framework to formalise suitable
features of selection rules has been highlighted. To this aim, a highly
expressive execution model based on priority mechanism for atom selection is
developed in the paper. The distinctive aspect is that primary importance is
given to the event of arrival of new atoms from the body of the applied
clause at rewriting time, when new atoms can be freely positioned with
respect to old ones in the resolvent. Then, at any derivation step, the atom
with optimum priority is simply selected.

The results presented in the paper show that the new computational model is
able to give remarkable insights into general properties of selection rules.
As a matter of fact, the priority model allows us to formalise the delicate
concepts on which the axiomatic definition of specialisation independent
scheduling rules is based. As a quite unexpected result, the specialisation
independence turns out to be equivalent to stack-queue scheduling technique,
which has a very simple operational characterisation. In other words, the
priority mechanism is necessary to formalise the real semantic features of
specialisation independent scheduling rules. On the contrary, the full
generality of the same mechanism can be abandoned if only operational
aspects of specialisation independent rules are of interest, in the sense
that all we need is a ``watershed'' between the stacked and the queued atoms.

It is widely acknowledged that the study of selection rules is a difficult
subject which deserves attention. We are confident that the computational
model proposed in the paper can be usefully exploited in future work to get
further insights into topics which are related to selection rule theory and
application, such as loop check, termination and optimisation of derivation
processes.

\appendix

\section{Appendix}

\noindent This Appendix contains the formal proofs of Properties \ref
{py2.5.1} and \ref{py3.1.1}. The very simple Property A1 is considered
before proving Property \ref{py2.5.1}.\bigskip

\begin{property}
\label{pyA1}Let$\;S\;$be a complete set of derivation steps. Given a p-goal$%
\;G\;$and a clause $c$, the following implication holds:\smallskip 

$\exists Ds$\ derivation step of the type $(G\overset{c\xi }{\longrightarrow 
}\bullet ),\hfill $(1)\qquad \smallskip 

$\qquad \Longrightarrow \;\exists Ds^{\prime }$\ of the type $(G\overset{%
c\xi }{\longrightarrow }\bullet )$, \ \ with $Ds^{\prime }\in S$.
\end{property}
\smallskip 

\begin{proof}
Let$\;G=a|K$ and $c=(ht\longleftarrow B)$. By (1)
and the completeness of$\;S\;$(part i), a derivation step exists of the
form:\smallskip

$(a|K\overset{c}{\longrightarrow }(K+B\xi ^{\prime }\underline{\theta })\mu
^{\prime })\in S\hfill $(2)\qquad \smallskip

\noindent By definition, the derivation step in (1) has the form:\smallskip

$a|K\overset{c}{\longrightarrow }(K+B\xi \underline{\tau })\mu .$\smallskip

\noindent Then, it is evident that a derivation also exists like:\smallskip

$Ds^{\prime }=(a|K\overset{c}{\longrightarrow }(K+B\xi \underline{\theta }%
)\mu ).$\smallskip

\noindent By construction, derivation steps (2) and $Ds^{\prime }$ are
congruent lowerings of each other. Then, by completeness of$\;S\;$(part ii),
derivation step $Ds^{\prime }$ belongs to$\;S$.
\end{proof}
\bigskip

\begin{property}[Property \ref{py2.5.1}]
Let$\;S\;$be a complete set of
derivation steps. Given two p-goals $a\gamma \underline{\tau }|G$\ and $a|F$%
, let us fix arbitrarily a finite set $V\;$of variables. The following
implication holds:

$\exists Ds$\ \ derivation step of the form $a\gamma \underline{\tau }|G%
\overset{c}{\longrightarrow }\bullet \hfill $(1)\qquad \smallskip 

$\;\;\;\Longrightarrow \exists Ds^{\prime }$\ \ of the form $a|F\overset{c}{%
\longrightarrow }\bullet $, \ with\ $Ds^{\prime }\in S\;$and $%
nvar(Ds^{\prime })\cap V=\varnothing .$\smallskip 
\end{property}

\begin{proof}
Let$\;c\;=(ht\longleftarrow B)$. On the basis of
(1), by definition of derivation step, a standardisation apart renaming $\xi
^{\prime }$ for$\;c\;$and an mgu $\beta $ exist, with $a\gamma \beta
=(ht)\xi ^{\prime }\beta $. Then, let us consider a renaming $\xi $ of $c\xi
^{\prime }$, such that the following assertions hold for the range of $\xi $%
:\smallskip

$var(a|F)\cap var(c\xi ^{\prime }\xi )=\varnothing ,$\hfill (2a)\qquad
\smallskip

$domain(\gamma )\cap var((ht)\xi ^{\prime }\xi )=\varnothing ,$\hfill
(2b)\qquad \smallskip

$domain(\xi ^{-1})\cap var(a\gamma )=\varnothing $\hfill (2c)\qquad
\smallskip

$var(c\xi ^{\prime }\xi )\cap V=\varnothing .$\hfill (2d)\qquad \smallskip

\noindent By facts (2b) and (2c), we have that:\smallskip

$a\gamma \xi ^{-1}\beta =^{(2c)}a\gamma \beta =(ht)\xi ^{\prime }\beta
=(ht)\xi ^{\prime }\xi \xi ^{-1}\beta =^{(2b)}(ht)\xi ^{\prime }\xi \gamma
\xi ^{-1}\beta .\smallskip $

\noindent In other words, $a$ and $(ht)\xi ^{\prime }\xi $ unify through the
unifier $\gamma \xi ^{-1}\beta $. On the other hand, the fact (2a) says that 
$\xi ^{\prime }\xi $ is a standardisation apart renaming for$\;c\;$with
respect to $a|F$. Then, a derivation step exists of the form $a|F\overset{%
c\xi ^{\prime }\xi }{\longrightarrow }\bullet $. By hypothesis the set$\;S\;$%
is complete, so that by Property A1 we have also a derivation step such
that:\smallskip

$Ds^{\prime }=(a|F\overset{c\xi ^{\prime }\xi }{\longrightarrow }\bullet
)\in S.$\smallskip

\noindent Since it is $nvar(Ds^{\prime })=var(c\xi ^{\prime }\xi )$, by (2d)
we have that $nvar(Ds^{\prime })\cap V=\varnothing $.
\end{proof}
\bigskip

\begin{property}[Property \ref{py3.1.1}]
Let$\;c\;=(ht\longleftarrow B)$\ be a
clause. Let us consider two derivation steps $Ds_{1}$\ and $Ds_{2}$\ such
that the $Ds_{2}$\ is a lowering of $Ds_{1}$\ by $X$\emph{.} The following
implication holds:

$Ds_{1}=(a|K\overset{c}{\longrightarrow }(K+B\xi ^{\prime }\underline{\theta 
}^{\prime })\mu ^{\prime }),$\hfill (1)\qquad \emph{\smallskip }

$Ds_{2}=(a\tau \underline{\sigma }|(K\tau \underline{\sigma }+X)\overset{c}{%
\longrightarrow }(K\tau \underline{\sigma }+B\xi ^{\prime \prime }\underline{%
\theta }^{\prime \prime }+X)\mu ^{\prime \prime })$\hfill (2)\qquad \emph{%
\smallskip }

$\qquad \Longrightarrow \;\;\exists \delta \;\;$such that \ $K\tau \mu
^{\prime \prime }=K\mu ^{\prime }\delta ,\;\;B\xi ^{\prime \prime }\mu
^{\prime \prime }=B\xi ^{\prime }\mu ^{\prime }\delta ,$

\qquad \qquad\ \ where $\delta $\ is a renaming, if $\tau $\ is a
renaming. 
\end{property}

\begin{proof}
By definition of derivation step, we have:\smallskip

$var(a|K)\cap var((ht\longleftarrow B)\xi ^{\prime })=\varnothing ,\hfill $%
(3)\qquad \smallskip

$var((a|K)\tau )\cap var((ht\longleftarrow B)\xi ^{\prime \prime
})=\varnothing $,\hfill (4)\qquad \smallskip

$\mu ^{\prime }=mgu(a,(ht)\xi ^{\prime }),\;\;\;\mu ^{\prime \prime
}=mgu(a\tau ,(ht)\xi ^{\prime \prime })$.\hfill (5)\qquad \emph{\smallskip }

\noindent Let $\pi =\tau /var(a|K)^{[}$\footnote{%
The notation $\tau /var(a|K)$ represents $\tau $ restricted to the variables
of $a|K$.}$^{]}$ \ \ and $\ \ \phi =((\xi ^{\prime })^{-1}\xi ^{\prime
\prime })/var((ht\longleftarrow B)\xi ^{\prime })$. By (3) it is:\smallskip

$domain(\pi )\cap domain(\phi )=\varnothing $,\hfill (6a)\qquad \smallskip

$(ht\longleftarrow B)\xi ^{\prime }\pi =(ht\longleftarrow B)\xi ^{\prime }$
and $(a|K)\phi =(a|K)$.\hfill (6b)\qquad \emph{\smallskip }

\noindent As a consequence of (6a), the union $(\pi \cup \phi )$ is a well
defined substitution. Then, we may write that:\smallskip

$a(\pi \cup \phi )\mu ^{\prime \prime }=^{(6b)}a\pi \mu ^{\prime \prime
}=a\tau \mu ^{\prime \prime }=^{(5)}(ht)\xi ^{\prime \prime }\mu ^{\prime
\prime }=(ht)\xi ^{\prime }(\xi ^{\prime })^{-1}\xi ^{\prime \prime }\mu
^{\prime \prime }=$\smallskip

$\hfill =(ht)\xi ^{\prime }\phi \mu ^{\prime \prime }=^{(6b)}(ht)\xi
^{\prime }(\pi \cup \phi )\mu ^{\prime \prime },\qquad $\smallskip

\noindent so that $(\pi \cup \phi )\mu ^{\prime \prime }$ is an unifier of $%
a $ and $(ht)\xi ^{\prime }$. Since $\mu ^{\prime }$ is an mgu of $a$ and $%
(ht)\xi ^{\prime }$, a substitution $\delta $ exists with:\smallskip

$(\pi \cup \phi )\mu ^{\prime \prime }=\mu ^{\prime }\delta $.\hfill
(7)\qquad \smallskip

\noindent Then, we have:\smallskip

$K\tau \mu ^{\prime \prime }=K\pi \mu ^{\prime \prime }=^{(6b)}K(\pi \cup
\phi )\mu ^{\prime \prime }=^{(7)}K\mu ^{\prime }\delta ,$\hfill (8a) \ \ \
\smallskip

$B\xi ^{\prime \prime }\mu ^{\prime \prime }=B\xi ^{\prime }(\xi ^{\prime
})^{-1}\xi ^{\prime \prime }\mu ^{\prime \prime }=B\xi ^{\prime }\phi \mu
^{\prime \prime }=^{(6b)}B\xi ^{\prime }(\pi \cup \phi )\mu ^{\prime \prime
}=^{(7)}B\xi ^{\prime }\mu ^{\prime }\delta .$\hfill (8b) \ \ \ \smallskip

\noindent Now let us suppose that $\tau $ is a renaming. In this case, facts
(3) and (4) become symmetric at all. As a consequence, by symmetry with
respect to (8a) and (8b), a substitution $\gamma $ exists such that $K\mu
^{\prime }=K\tau \mu ^{\prime \prime }\gamma $ and $B\xi ^{\prime }\mu
^{\prime }=B\xi ^{\prime \prime }\mu ^{\prime \prime }\gamma $. Then we
have:\smallskip

$(K\mu ^{\prime }+B\xi ^{\prime }\mu ^{\prime })\delta \gamma =(K\tau \mu
^{\prime \prime }+B\xi ^{\prime \prime }\mu ^{\prime \prime })\gamma =K\mu
^{\prime }+B\xi ^{\prime }\mu ^{\prime }.$\smallskip

\noindent It is evident that $\delta $ is a renaming for $K\mu ^{\prime
}+B\xi ^{\prime }\mu ^{\prime }$, then the thesis is verified.
\end{proof}

\section{Appendix}

In this Appendix we provide a formal proof of the duplication theorem
(Theorem \ref{t4.3.1}). Such a proof exploits two lemmata which are given
below. Lemma B1 establishes a condition which allows us to repeat
derivations via a specialisation independent scheduling rule, when we pass
from a goal$\;G\;$to a suitable kind of instantiations of $G$. Lemma B1 is a
correspondent, for p-SLD derivations, of part (ii) of Strong Lifting Lemma 
\cite{7.GLM96}. Indeed, both part (ii) of Strong Lifting Lemma and Lemma B1
can be seen as results about sufficient conditions for derivation lowering
from a goal$\;G\;$to instantiations of$\;G\;$itself. Here a direct proof of
Lemma B1 is given which takes into account technical aspects concerning our
priority value mechanism. Lemma B1 does not relate resolvents, because it is
not important for the purposes of this Appendix.\medskip

\begin{lemma}
\label{LB1}Let$\;S\;$be a specialisation independent scheduling rule,$\;G\;$%
a p-goal and $\phi $\ a substitution. The following implication
holds:\smallskip 

$G\overset{S,X,\theta }{\longrightarrow }\bullet \;\;\;\Longrightarrow
\;\;\;G\theta \phi \overset{S,X}{\longrightarrow }\bullet .\hfill $(1)\qquad
\smallskip 
\end{lemma}

\begin{proof}
The proof is by induction on the length of $X$. If 
$\#X=0$, the thesis is trivially true. For $\#X>0$, let$\;G=a|F$, $X=c|H$
with$\;c=(ht\longleftarrow B)$, and rewrite derivation (1) as
follows:\smallskip

$a|F\overset{S,c\xi ,\gamma }{\longrightarrow }(Q=(F+B\xi \underline{\pi }%
)\gamma )\overset{S,H,\mu }{\longrightarrow }\bullet ,$ \ \ where $\gamma
\mu =\theta $.\hfill (2)\qquad \smallskip

\noindent Then, let us consider the substitution $\phi _{g}=\phi \sigma
_{g}, $ where $\sigma _{g}$ is such that $(a|F)\theta \phi _{g}=G\theta \phi
_{g}$ is ground. Since $\gamma $ is an mgu of $a$ and $(ht)\xi $, we have $%
a\gamma =(ht)\xi \gamma $, which means $a\theta \phi _{g}=a\gamma \mu \phi
_{g}=(ht)\xi \gamma \mu \phi _{g}=(ht)\xi \theta \phi _{g}.$ But $a\theta
\phi _{g}$ is ground, so that we obtain the equality $(a\theta \phi
_{g})\theta \phi _{g}=a\theta \phi _{g}=((ht)\xi )\theta \phi _{g}$. In
other words, $a\theta \phi _{g}$ and $(ht)\xi $ unify through the unifier $%
\theta \phi _{g}$. Moreover, the renamed clause $c\xi $ is obviously
standardised apart with respect to the ground p-goal $(a|F)\theta \phi _{g}$%
, so that a derivation step like $(a|F)\theta \phi _{g}\overset{c\xi }{%
\longrightarrow }\bullet $ exists. Thus, by completeness of$\;S\;$and
Property A1, a derivation step also exists of the form:\smallskip

$(G\theta \phi _{g}=(a|F)\theta \phi _{g})\overset{S,c\xi ,\eta }{%
\longrightarrow }(R=(F\theta \phi _{g}+B\xi \underline{\pi }^{\prime })\eta
).$\hfill (3)\qquad \smallskip

\noindent Now, the substitution $\eta $ is an mgu of $a\theta \phi _{g}$ and 
$(ht)\xi $, so that a substitution $\pi $ exists with:\smallskip

$\theta \phi _{g}=\eta \pi .$\hfill (4)\qquad \smallskip

\noindent On the other hand, since$\;S\;$is specialisation independent, step
(3) is a congruent lowering of the first step of (2) by $\varnothing $,
i.e.\smallskip

$\exists \underline{\rho }$ such that $F\underline{\rho }=F,\;B\underline{%
\pi }\underline{\rho }=B\underline{\pi }^{\prime }$,\hfill (5)\qquad
\smallskip

\noindent which implies:\emph{\medskip }

$Q\mu \phi _{g}=(F+B\xi \underline{\pi })\gamma \mu \phi _{g}\underline{\rho 
}\underline{\rho }^{-1}=(F\underline{\rho }+B\xi \underline{\pi }\underline{%
\rho })\theta \phi _{g}\underline{\rho }^{-1}=^{(5)}(F+B\xi \underline{\pi }%
^{\prime })\theta \phi _{g}\underline{\rho }^{-1}$.\emph{\medskip }

\noindent But $F\theta \phi _{g}$ is ground, so that $(F\theta \phi
_{g})\theta \phi _{g}=F\theta \phi _{g}$. As a consequence:\emph{\medskip }

$Q\mu \phi _{g}=(F\theta \phi _{g}+B\xi \underline{\pi }^{\prime })\theta
\phi _{g}\underline{\rho }^{-1}=^{(4)}(F\theta \phi _{g}+B\xi \underline{\pi 
}^{\prime })\eta \pi \underline{\rho }^{-1}=R\pi \underline{\rho }^{-1}.$%
\emph{\medskip }

\noindent By inductive hypothesis applied to the tail of (2), we have that $%
(Q\mu \phi _{g}=R\pi \underline{\rho }^{-1})\overset{S,H}{\longrightarrow }%
\bullet ,$ which by Lifting Lemma \ref{L3.2.1} implies that$\;R\overset{S,H}{%
\longrightarrow }\bullet $. Now, by Property \ref{py3.2.1}, the last
obtained derivation can be combined with (3) yielding:\smallskip

$(G\theta \phi _{g}=G\theta \phi \sigma _{g})\overset{S,X}{\longrightarrow }%
\bullet .$\smallskip

\noindent By Lifting Lemma \ref{L3.2.1}, we conclude $(G\theta \phi \overset{%
S,X}{\longrightarrow }\bullet )$, so that the inductive step is completed.
\end{proof}
\medskip

The following Lemma B2 is a special form of determinism lemma which holds
for preq type stack-queue derivations. Roughly speaking, the lemma states
that an $A$-preq type derivation, starting from a p-goal of the form $A|X$,
can be replicated from a p-goal like $A\lambda \underline{\lambda }|Y$,
where $\lambda $ is a renaming. Note that no hypothesis is made on $X$ and $%
Y $ which can be completely unrelated. The intuitive explication is that
only atoms deriving from $A$ are rewritten so that neither $X$ nor $Y$ have
any active role in the derivations. The formal statement and the proof of
Lemma B2 are preceded by the quite simple Property B1.
\medskip

\begin{property}
\label{pyB1}Let $SQ$\ be a stack-queue scheduling rule. The following
implication holds:\emph{\smallskip }

$A|X\overset{SQ,D}{\longrightarrow }Q$, of $A$-preq type\emph{\hfill }%
(1)\qquad \emph{\smallskip }

$A\gamma \underline{\lambda }|Y\overset{SQ,D}{\longrightarrow }R$, of $%
(A\gamma \underline{\lambda })$-preq type, where $\gamma $ is a renaming%
\emph{\hfill }(2)\qquad \emph{\smallskip }

$\Longrightarrow $ $\exists \delta ,\underline{\delta }$\ such that $%
R/(A\gamma \underline{\lambda })=(Q/A)\delta \underline{\delta },$\ where $%
\delta $\ is a renaming.\smallskip 
\end{property}

\begin{proof}
By hypothesis, derivations (1) and (2) are of $A$%
-preq and $(A\gamma \underline{\lambda })$-preq type respectively, so that $%
D/A=D/(A\gamma \underline{\lambda })=D$. Then, by Lifting Lemma \ref{L3.2.1}%
, a derivation exists like:\smallskip

$A\overset{S,D}{\longrightarrow }T$.\hfill (3)\qquad \smallskip

\noindent By Lowering Lemma \ref{L3.1.1}, applied to (3) and (1), a renaming 
$\alpha $ and a shifting $\underline{\alpha }$ exist with $Q/A=T\alpha 
\underline{\alpha }$. By Lowering Lemma \ref{L3.1.1} applied to (3) and (2),
a renaming $\beta $ and a shifting $\underline{\beta }$ exist with $%
R/(A\gamma \underline{\lambda })=T\beta \underline{\beta }$. Finally, we
derive that:\smallskip

$R/(A\gamma \underline{\lambda })=T\alpha \underline{\alpha }\alpha ^{-1}%
\underline{\alpha }^{-1}\beta \underline{\beta }=(Q/A)\alpha ^{-1}\underline{%
\alpha }^{-1}\beta \underline{\beta }$.
\end{proof}
\medskip

\begin{lemma}[preq type determinism]
\label{LB2}
Let $SQ$\ be a stack-queue scheduling
rule and $V$\ any finite set of variables. Let $A|X$\ and $A\lambda 
\underline{\lambda }|Y$\ two p-goals, where $\lambda $\ is a renaming. The
following implication holds:\medskip 

$A|X\overset{SQ,K,\psi }{\longrightarrow }A^{s}|X\psi |A^{q}$, of $A$-preq
type,\hfill (1)\qquad \smallskip 

$\Longrightarrow $ $\exists \delta ,\underline{\delta }$\ and $D=(A\lambda 
\underline{\lambda }|Y\overset{SQ,K,\theta }{\longrightarrow }A^{s}\delta 
\underline{\delta }|Y\theta |A^{q}\delta \underline{\delta })$, of $%
(A\lambda \underline{\lambda })$-preq type,\smallskip 

\qquad where $\delta $\ is a renaming and $nvar(D)\cap V=\varnothing .$
\end{lemma}

\begin{proof}
Let $A=a|F$. We show Lemma B2 by induction on the
length of the template $K$. If $\#K=0$, the assert is evident. If $\#K>0$,
let $K=c|H$ with$\;c=(ht\longleftarrow M_{s}|M_{q})$. Derivation (1) can be
rewritten as follows:\smallskip

$a|F|X\overset{SQ,c\xi ,\mu }{\longrightarrow }(Q=a^{s}|F\mu |X\mu |a^{q})%
\overset{SQ,H,\sigma }{\longrightarrow }A^{s}|X\psi |A^{q}$\hfill (1a)\qquad
\smallskip

with $a^{s}=M_{s}\xi \underline{\alpha }\mu $ \ and $\ A^{s}=(a^{s}|F\mu
)^{s}$.\hfill (1b)\qquad \smallskip

\noindent Then, by Property \ref{py2.5.1} with reference to the first step
of (1a), a derivation step $Ds$ exists such that:\smallskip

$Ds=((a|F)\lambda \underline{\lambda }|Y\overset{SQ,c\tau ,\eta }{%
\longrightarrow }R)$, \ with $nvar(Ds)\cap V=\varnothing $.\hfill (2)\qquad
\smallskip

\noindent Since the selection rule $SQ$ is stack queue, it must be:\smallskip

$R=M_{s}\tau \underline{\gamma }\eta |F\lambda \underline{\lambda }\eta
|Y\eta |M_{q}\tau \underline{\gamma }\eta $.\smallskip

\noindent By Property B1 applied to derivation step (2) and the first step
of (1a), a renaming $\beta $ and a shifting $\underline{\beta }$ exist with:%
\emph{\medskip }

$M_{s}\tau \underline{\gamma }\eta |F\lambda \underline{\lambda }\eta
|M_{q}\tau \underline{\gamma }\eta =R/^{2}/(A\lambda \underline{\lambda }%
)=(Q/^{1a}/A)\beta \underline{\beta }=(a^{s}|F\mu |a^{q})\beta \underline{%
\beta }.$\emph{\medskip }

\noindent Now, by (1b) it is $\ \#a^{s}=\#M_{s}\tau \underline{\gamma }\eta
, $ \ so that the equality $\;M_{s}\tau \underline{\gamma }\eta |F\lambda 
\underline{\lambda }\eta =(a^{s}|F\mu )\beta \underline{\beta }$ \ holds, by
Property \ref{py2.1.1}-i). Thus, the inductive hypothesis can be applied to
the tail of (1a). As a consequence, a p-SLD derivation $D^{\prime }$ exists,
which is of $(M_{s}\tau \underline{\gamma }\eta |F\lambda \underline{\lambda 
}\eta )$-preq type and has the following form:\emph{\medskip }

$R\overset{SQ,H,\pi }{\longrightarrow }((a^{s}|F\mu )^{s}\delta ^{\prime }%
\underline{\delta }^{\prime }|Y\eta \pi |Z=^{(1b)}A^{s}\delta ^{\prime }%
\underline{\delta }^{\prime }|Y\eta \pi |Z)$,\hfill (3)\qquad \smallskip

with $nvar(D^{\prime })\cap (var((a|F)\lambda \underline{\lambda }|Y)\cup
nvar(Ds)\cup V)=\varnothing $.\hfill (3a)\qquad \emph{\medskip }

\noindent On the basis of (3a) above, derivation step (2) and derivation (3)
can be combined to yield the derivation $D$:\emph{\medskip }

$D=(A\lambda \underline{\lambda }|Y\overset{SQ,K}{\longrightarrow }%
A^{s}\delta ^{\prime }\underline{\delta }^{\prime }|Y\eta \pi |Z)$,\hfill
(4)\qquad \emph{\medskip }

\noindent where $D$ is of $(A\lambda \underline{\lambda })$-preq type. The
thesis is now proven. Indeed, by Property B1 applied to derivations (1) and
(4), a renaming $\delta $ and a shifting $\underline{\delta }$ exist with $%
A^{s}\delta ^{\prime }\underline{\delta }^{\prime }|Z=(A^{s}|A^{q})\delta 
\underline{\delta }$, so that by Property \ref{py2.1.1}-i) we have $%
A^{s}\delta ^{\prime }\underline{\delta }^{\prime }=A^{s}\delta \underline{%
\delta }$ and $Z=A^{q}\delta \underline{\delta }$. The fact that $%
nvar(D)\cap V=\varnothing $ follows from (2) and (3a).
\end{proof}
\bigskip

\begin{theorem}[Theorem \ref{t4.3.1} -- duplication theorem]
Let $SQ$ be a
stack-queue scheduling rule. Given two p-goals of the form $A|B|C|D$\ and $%
A|B|C|B\underline{\pi }|D$, the following implication holds:\smallskip

$(A|B|C|D\overset{SQ,X.P}{\longrightarrow }Q)\hfill $(1)\qquad \smallskip

$\Longrightarrow $ $\exists Y$\ such that $(A|B|C|B\underline{\pi }|D%
\overset{SQ,Y.P}{\longrightarrow }R)$\smallskip

\qquad with $X\subseteq _{L}Y$\ and $\#Q\preceq \#R.$
\end{theorem}

\begin{proof}
Let $\Delta (SQ,n)$ denote the subset of $\Delta
(SQ)$, such that for any derivation $Dr$ in $\Delta (SQ,n)$ it is $%
\#Dr\preceq n$, where $\#Dr$ denotes the length of $Dr$. We show the thesis
by induction on $n$, i.e. we show that the thesis holds when derivation (1)
belongs to $\Delta (SQ,n)$, for any $n\succeq 0$. The fact is obvious for $%
\Delta (SQ,0)$. In order to prove the inductive step from $\Delta (SQ,n-1)$
to $\Delta (SQ,n)$, for $n>0$, let us consider a derivation like:\smallskip

$(A|B|C|D\overset{X.P,\theta }{\longrightarrow }Q)\in \Delta (SQ,n)$,\hfill
(1a)\qquad \smallskip

\noindent and show that $(A|B|C|B\underline{\pi }|D\overset{SQ,Y.P}{%
\longrightarrow }R)$ exists with $X\subseteq _{L}Y$ and $\#Q\preceq \#R$.
Actually, the proof of the inductive step will be organised in two phases:

\begin{itemize}
\item  first, the inductive step is shown in the case that the initial
p-goal $A|B|C|D$ is ground,

\item  then, the validity of the inductive step is extended to generic
initial p-goals.
\end{itemize}

\noindent Let us recall that the sketch of Section \ref{subs4.3} was given
in the simplifying hypothesis that: every clause body introduces no new
variable and initial p-goals are ground. In this sense, we may say that the
first phase removes the first restriction, while the second one is retained.
In the second phase, also the restriction on the groundness of initial goals
is overcome.\bigskip

\noindent \underline{First phase}\emph{\ (the initial p-goal }$A|B|C|D$\emph{%
\ is ground).\medskip }

\noindent With reference to (1a), the following three possible situations
must be taken into account. Then, we start with Case 3, which is the most
significant one.

\begin{enumerate}
\item  derivation (1a) is of $(A|B|C)$-preq type,\medskip

\item  derivation (1a) is of $(A|B|C|D)$-preq type, and not of $(A|B|C)$%
-preq type,\medskip

\item  derivation (1a) is not of $(A|B|C|D)$-preq type.\medskip
\end{enumerate}

\noindent \underline{Case3}.\smallskip

\noindent Derivation (1a) has the following form:\smallskip

$A|B|C|D\overset{H}{\longrightarrow }B|C|D|A^{q}\overset{K}{\longrightarrow }%
C|D|A^{q}|B^{q}\overset{M}{\longrightarrow }$\smallskip

\qquad $D|A^{q}|B^{q}|C^{q}\overset{N}{\longrightarrow }%
A^{q}|B^{q}|C^{q}|D^{q}\overset{T}{\longrightarrow }Q,$\hfill (2)\qquad
\smallskip

where $H|K|M|N|T=X$.

\noindent In fact, since $A|B|C|D$ is ground, in each of the four initial
segments of (2) only standardisation apart variables, introduced in the same
segment, can be instantiated. In particular, $var(A^{q})$ are not
instantiated in the second segment, $var(A^{q}|B^{q})$ are not in the third
segment, and $var(A^{q}|B^{q}|C^{q})$ are not in the fourth one. It is
evident, as a consequence, that the p-goal $A^{q}|B^{q}|C^{q}|D^{q}$
consists of four subgoals without common variables. Now, since also $B%
\underline{\pi }$ is ground, a derivation can be constructed through five
successive applications of Lemma B2, as depicted below:\smallskip

$A|B|C|B\underline{\pi }|D\overset{SQ,H}{\longrightarrow }B|C|B\underline{%
\pi }|D|A^{q}\alpha \underline{\alpha }\overset{SQ,K}{\longrightarrow }$%
\smallskip

\qquad $C|B\underline{\pi }|D|A^{q}\alpha \underline{\alpha }|B^{q}\beta 
\underline{\beta }\overset{SQ,M}{\longrightarrow }B\underline{\pi }%
|D|A^{q}\alpha \underline{\alpha }|B^{q}\beta \underline{\beta }|C^{q}\gamma 
\underline{\gamma }\overset{SQ,K}{\longrightarrow }$\hfill (3)\qquad
\smallskip

\qquad $D|A^{q}\alpha \underline{\alpha }|B^{q}\beta \underline{\beta }%
|C^{q}\gamma \underline{\gamma }|B^{q}\phi \underline{\phi }\overset{SQ,N}{%
\longrightarrow }(Z=A^{q}\alpha \underline{\alpha }|B^{q}\beta \underline{%
\beta }|C^{q}\gamma \underline{\gamma }|B^{q}\phi \underline{\phi }%
|D^{q}\delta \underline{\delta })$\smallskip

where $\alpha $, $\beta $, $\gamma $, $\phi $ and $\delta $ are
renamings.\smallskip

\noindent At each application of Lemma B2, a segment of derivation (3) is
obtained on the basis of a corresponding segment of derivation (2).
Moreover, Lemma B2 assures that each new segment can be freely standardised
apart, so that each segment can be readily added to the sequence of its
predecessors in (3). Note that the second segment of (2) is considered
twice, in order to generate both the second and the fourth segment of (3).
In analogy with derivation (2), the final p-goal $Z$ of derivation (3)
consists of five subgoals without common variables. As a consequence, the
five renamings $\alpha ^{-1}$, $\beta ^{-1}$, $\gamma ^{-1}$, $\phi ^{-1}$
and $\delta ^{-1}$ have disjoint domains, so that they can be joined in
order to form a unique substitution\smallskip

$\xi =(\alpha ^{-1}\cup \beta ^{-1}\cup \gamma ^{-1}\cup \phi ^{-1}\cup
\delta ^{-1})$.\hfill (4a)\qquad \smallskip

\noindent Then, let us consider the p-goal $A^{q}|B^{q}|C^{q}|B^{q}%
\underline{\pi }^{\prime }|D^{q}$, where $\underline{\pi }^{\prime }$ is a
suitable shifting. By (4a) and Property \ref{py2.2.1}, we have
that:\smallskip

$A^{q}|B^{q}|C^{q}|B^{q}\underline{\pi }^{\prime }|D^{q}=^{(Ax-iii)}(A^{q}%
\underline{\alpha }|B^{q}\underline{\beta }|C^{q}\underline{\gamma }|B%
\underline{\phi }|D^{q}\underline{\delta })\underline{\sigma }=^{(4a)}$%
\smallskip

\qquad $(A^{q}\alpha \underline{\alpha }|B^{q}\beta \underline{\beta }%
|C^{q}\gamma \underline{\gamma }|B^{q}\phi \underline{\phi }|D^{q}\delta 
\underline{\delta })\xi \underline{\sigma }=Z\xi \underline{\sigma }$.\hfill
(4b)\qquad \smallskip

\noindent By construction of (2) the derivation $(A^{q}|B^{q}|C^{q}|D^{q}%
\overset{T}{\longrightarrow }Q)$ belongs to $\Delta (SQ,m)$, with $m<n$. By
inductive hypothesis, a derivation exists of the form:\smallskip

$(Z\xi \underline{\sigma }=A^{q}|B^{q}|C^{q}|B^{q}\underline{\pi }^{\prime
}|D^{q})\overset{SQ,Y^{\prime }.P}{\longrightarrow }R^{\prime }$,\hfill
(5)\qquad \smallskip

with $\ T\subseteq _{L}Y^{\prime }$ \ and $\#Q\preceq \#R^{\prime }$.\hfill
(5a)\qquad \smallskip

\noindent By (5) and Lifting Lemma \ref{L3.2.1} a derivation exists
like:\smallskip

$Z\overset{SQ,Y^{\prime }}{\longrightarrow }\bullet $,\hfill (6)\qquad
\smallskip

\noindent By Property \ref{py3.2.1}, derivations (3) and (6) can be combined
to yield a derivation of the form:\smallskip

$A|B|C|B\underline{\pi }|D\overset{SQ,(H|K|M|K|N).P}{\longrightarrow }Z%
\overset{SQ,Y^{\prime }.P}{\longrightarrow }R$,\hfill (7)\qquad \smallskip

\noindent where, by Lowering Lemma \ref{L3.1.1} applied to (5) and the tail
of (7), it is $\#R=\#R^{\prime }$. Finally:\smallskip

$X=H|K|M|N|T\subseteq _{L}^{(5a)}(H|K|M|K|N|Y^{\prime })$ \ and $\#Q\preceq
^{(5a)}\#R^{\prime }=\#R$.\medskip

\noindent \underline{Case 2}.\smallskip

\noindent Derivation (1a) has the following form:\smallskip

$A|B|C|D\overset{H|K|M|N}{\longrightarrow }D^{s}|A^{q}|B^{q}|C^{q}|D^{q}$,
with $H|K|M|N=X$.\smallskip

\noindent Analogously to preceding case 3), through Lemma B2 a derivation
can be constructed like:\smallskip

$A|B|C|B\underline{\pi }|D\overset{SQ,(H|K|M|K|N).P}{\longrightarrow }%
D^{s}\delta \underline{\delta }|A^{q}\alpha \underline{\alpha }|B^{q}\beta 
\underline{\beta }|C^{q}\gamma \underline{\gamma }|B^{q}\phi \underline{\phi 
}|D^{q}\delta \underline{\delta }$.\medskip

\noindent \underline{Case 1}.\smallskip

\noindent Derivation (1a) has the form:\smallskip

$A|B|C|D\overset{X}{\longrightarrow }(A|B|C)^{s}|D|(A|B|C)^{q}.$\smallskip

\noindent Through Lemma B2, a derivation can be constructed like:\smallskip

$(A|B|C)|B\underline{\pi }|D\overset{SQ,X}{\longrightarrow }%
(A|B|C)^{s}\gamma \underline{\gamma }|B\underline{\pi }|D|(A|B|C)^{q}\gamma 
\underline{\gamma }.$
\bigskip
\smallskip

\noindent \underline{Second phase}\emph{\ (the initial p-goal }$A|B|C|D$%
\emph{\ is generic).\medskip }

\noindent In the preceding first phase of this proof, the inductive step is
verified in the hypothesis that the initial p-goal $A|B|C|D$ is ground. Now
consider a generic p-goal of the form $A|B|C|D$. With reference to (1a), let 
$\phi _{g}$ be a grounding substitution for $(A|B|C|D)\theta $. By Lemma B1,
a derivation exists such that:\smallskip

$((A|B|C|D)\theta \phi _{g}\overset{X}{\longrightarrow }Q^{\prime })\in
\Delta (SQ,n)$,\hfill (8)\qquad \smallskip

\noindent where, by Lowering Lemma \ref{L3.1.1} applied to (1a) and (8), we
have:\smallskip

$\#Q^{\prime }=\#Q$.\hfill (8a)\qquad \smallskip

\noindent Since the inductive hypothesis is already proven for ground
initial goals, by (8) a derivation exists:\smallskip

$(A|B|C|B\underline{\pi }|D)\theta \phi _{g}\overset{SQ,Y.P}{\longrightarrow 
}R^{\prime }$,\hfill (9)\qquad \smallskip

with $X\subseteq _{L}Y$ and $\#Q^{\prime }\preceq \#R^{\prime }$.\hfill
(9a)\qquad \smallskip

\noindent Then, by Lifting Lemma \ref{L3.2.1} a derivation exists:\smallskip

$A|B|C|B\underline{\pi }|D\overset{SQ,Y.P}{\longrightarrow }R$,\hfill
(10)\qquad \smallskip

\noindent where, by Lowering Lemma \ref{L3.1.1} applied to (9) and (10), we
have\smallskip

$X\subseteq _{L}^{(9a)}Y$ and $\#Q=^{(8a)}\#Q^{\prime }\preceq
^{(9a)}\#R^{\prime }=^{(Lem.\ref{L3.1.1})}\#R$.\smallskip

\noindent As a consequence, the induction step is completely verified.
\end{proof}

\end{document}